
\input fontch.tex

%
%
%
\def\unredoffs{} \def\redoffs{\voffset=-.31truein\hoffset=-.48truein}
\def\speclscape{}
%
%
%
%
%
\newbox\leftpage \newdimen\fullhsize \newdimen\hstitle \newdimen\hsbody
\tolerance=1000\hfuzz=2pt
\catcode`\@=11 
\ifx\hyperdef\UNd@FiNeD\def\hyperdef#1#2#3#4{#4}\def\hyperref#1#2#3#4{#4}\fi
\def\bigans{b }
\def\answ{b }
%
\ifx\answ\bigans\message{(This will come out unreduced.}
\magnification=1200\unredoffs\baselineskip=16pt plus 2pt minus 1pt
\hsbody=\hsize \hstitle=\hsize 
\else\message{(This will be reduced.} \let\l@r=L
\magnification=1000\baselineskip=16pt plus 2pt minus 1pt \vsize=7truein
\redoffs \hstitle=8truein\hsbody=4.75truein\fullhsize=10truein\hsize=\hsbody
\output={\ifnum\pageno=0 
  \shipout\vbox{\speclscape{\hsize\fullhsize\makeheadline}
    \hbox to \fullhsize{\hfill\pagebody\hfill}}\advancepageno
  \else
  \almostshipout{\leftline{\vbox{\pagebody\makefootline}}}\advancepageno
  \fi}
\def\almostshipout#1{\if L\l@r \count1=1 \message{[\the\count0.\the\count1]}
      \global\setbox\leftpage=#1 \global\let\l@r=R
 \else \count1=2
  \shipout\vbox{\speclscape{\hsize\fullhsize\makeheadline}
      \hbox to\fullhsize{\box\leftpage\hfil#1}}  \global\let\l@r=L\fi}
\fi
%
\newcount\yearltd\yearltd=\year\advance\yearltd by -2000

\def\Title#1#2{\nopagenumbers\abstractfont\hsize=\hstitle\rightline{#1}%
\vskip 1in\centerline{\titlefont #2}\abstractfont\vskip .5in\pageno=0}
\def\Date#1{\vfill\leftline{#1}\tenpoint\supereject\global\hsize=\hsbody%
\footline={\hss\tenrm\hyperdef\hypernoname{page}\folio\folio\hss}}%
%

\def\draftmode{\message{ DRAFTMODE }\def\draftdate{{\rm preliminary draft:
\number\month/\number\day/\number\yearltd\ \ \hourmin}}%
\headline={\hfil\draftdate}\writelabels\baselineskip=20pt plus 2pt minus 2pt
 {\count255=\time\divide\count255 by 60 \xdef\hourmin{\number\count255}
  \multiply\count255 by-60\advance\count255 by\time
  \xdef\hourmin{\hourmin:\ifnum\count255<10 0\fi\the\count255}}}
\def\nolabels{\def\wrlabeL##1{}\def\eqlabeL##1{}\def\reflabeL##1{}}
\def\writelabels{\def\wrlabeL##1{\leavevmode\vadjust{\rlap{\smash%
{\line{{\escapechar=` \hfill\rlap{\sevenrm\hskip.03in\string##1}}}}}}}%
\def\eqlabeL##1{{\escapechar-1\rlap{\sevenrm\hskip.05in\string##1}}}%
\def\reflabeL##1{\noexpand\llap{\noexpand\sevenrm\string\string\string##1}}}
\nolabels
%
\global\newcount\secno \global\secno=0
\global\newcount\meqno \global\meqno=1
\def\s@csym{}
\def\newsec#1{\global\advance\secno by1%
{\toks0{#1}\message{(\the\secno. \the\toks0)}}%
\global\subsecno=0\eqnres@t\let\s@csym\secsym\xdef\secn@m{\the\secno}\noindent
{\bf\hyperdef\hypernoname{section}{\the\secno}{\the\secno.} #1}%
\writetoca{{\string\hyperref{}{section}{\the\secno}{\the\secno.}} {#1}}%
\par\nobreak\medskip\nobreak}
\def\eqnres@t{\xdef\secsym{\the\secno.}\global\meqno=1\bigbreak\bigskip}
\def\sequentialequations{\def\eqnres@t{\bigbreak}}\xdef\secsym{}
\global\newcount\subsecno \global\subsecno=0
\def\subsec#1{\global\advance\subsecno by1%
{\toks0{#1}\message{(\s@csym\the\subsecno. \the\toks0)}}%
\ifnum\lastpenalty>9000\else\bigbreak\fi
\noindent{\it\hyperdef\hypernoname{subsection}{\secn@m.\the\subsecno}%
{\secn@m.\the\subsecno.} #1}\writetoca{\string\quad
{\string\hyperref{}{subsection}{\secn@m.\the\subsecno}{\secn@m.\the\subsecno.}}
{#1}}\par\nobreak\medskip\nobreak}
\def\appendix#1#2{\global\meqno=1\global\subsecno=0\xdef\secsym{\hbox{#1.}}%
\bigbreak\bigskip\noindent{\bf Appendix \hyperdef\hypernoname{appendix}{#1}%
{#1.} #2}{\toks0{(#1. #2)}\message{\the\toks0}}%
\xdef\s@csym{#1.}\xdef\secn@m{#1}%
\writetoca{\string\hyperref{}{appendix}{#1}{Appendix {#1.}} {#2}}%
\par\nobreak\medskip\nobreak}
%
%
\def\checkm@de#1#2{\ifmmode{\def\f@rst##1{##1}\hyperdef\hypernoname{equation}%
{#1}{#2}}\else\hyperref{}{equation}{#1}{#2}\fi}
\def\eqnn#1{\DefWarn#1\xdef #1{(\noexpand\relax\noexpand\checkm@de%
{\s@csym\the\meqno}{\secsym\the\meqno})}%
\wrlabeL#1\writedef{#1\leftbracket#1}\global\advance\meqno by1}
\def\f@rst#1{\c@t#1a\em@ark}\def\c@t#1#2\em@ark{#1}
\def\eqna#1{\DefWarn#1\wrlabeL{#1$\{\}$}%
\xdef #1##1{(\noexpand\relax\noexpand\checkm@de%
{\s@csym\the\meqno\noexpand\f@rst{##1}}{\hbox{$\secsym\the\meqno##1$}})}
\writedef{#1\numbersign1\leftbracket#1{\numbersign1}}\global\advance\meqno by1}
\def\eqn#1#2{\DefWarn#1%
\xdef #1{(\noexpand\hyperref{}{equation}{\s@csym\the\meqno}%
{\secsym\the\meqno})}$$#2\eqno(\hyperdef\hypernoname{equation}%
{\s@csym\the\meqno}{\secsym\the\meqno})\eqlabeL#1$$%
\writedef{#1\leftbracket#1}\global\advance\meqno by1}
\def\xeqn{\expandafter\xe@n}\def\xe@n(#1){#1}
\def\xeqna#1{\expandafter\xe@n#1}
\def\eqns#1{(\e@ns #1{\hbox{}})}
\def\e@ns#1{\ifx\UNd@FiNeD#1\message{eqnlabel \string#1 is undefined.}%
\xdef#1{(?.?)}\fi{\let\hyperref=\relax\xdef\next{#1}}%
\ifx\next\em@rk\def\next{}\else%
\ifx\next#1\xeqn#1\else\def\n@xt{#1}\ifx\n@xt\next#1\else\xeqna#1\fi
\fi\let\next=\e@ns\fi\next}

\def\DefWarn#1{\ifx\UNd@FiNeD#1\else
\immediate\write16{*** WARNING: the label \string#1 is already defined ***}\fi}
%
\newskip\footskip\footskip14pt plus 1pt minus 1pt 
\def\footnotefont{\ninepoint}\def\f@t#1{\footnotefont #1\@foot}
\def\f@@t{\baselineskip\footskip\bgroup\footnotefont\aftergroup\@foot\let\next}
\setbox\strutbox=\hbox{\vrule height9.5pt depth4.5pt width0pt}
\global\newcount\ftno \global\ftno=0
\def\foot{\global\advance\ftno by1\def\foot@rg{\hyperref{}{footnote}%
{\the\ftno}{\the\ftno}\xdef\foot@rg{\noexpand\hyperdef\noexpand\hypernoname%
{footnote}{\the\ftno}{\the\ftno}}}\footnote{$^{\foot@rg}$}}
%
\newwrite\ftfile
\def\footend{\def\foot{\global\advance\ftno by1\chardef\wfile=\ftfile
\hyperref{}{footnote}{\the\ftno}{$^{\the\ftno}$}%
\ifnum\ftno=1\immediate\openout\ftfile=\jobname.fts\fi%
\immediate\write\ftfile{\noexpand\smallskip%
\noexpand\item{\noexpand\hyperdef\noexpand\hypernoname{footnote}
{\the\ftno}{f\the\ftno}:\ }\pctsign}\findarg}%
\def\footatend{\vfill\eject\immediate\closeout\ftfile{\parindent=20pt
\centerline{\bf Footnotes}\nobreak\bigskip\input \jobname.fts }}}
\def\footatend{}
%
%
\global\newcount\refno \global\refno=1
\newwrite\rfile
\def\ref{[\hyperref{}{reference}{\the\refno}{\the\refno}]\nref}
\def\nref#1{\DefWarn#1%
\xdef#1{[\noexpand\hyperref{}{reference}{\the\refno}{\the\refno}]}%
\writedef{#1\leftbracket#1}%
\ifnum\refno=1\immediate\openout\rfile=\jobname.refs\fi
\chardef\wfile=\rfile\immediate\write\rfile{\noexpand\item{[\noexpand\hyperdef%
\noexpand\hypernoname{reference}{\the\refno}{\the\refno}]\ }%
\reflabeL{#1\hskip.31in}\pctsign}\global\advance\refno by1\findarg}
\def\findarg#1#{\begingroup\obeylines\newlinechar=`\^^M\pass@rg}
{\obeylines\gdef\pass@rg#1{\writ@line\relax #1^^M\hbox{}^^M}%
\gdef\writ@line#1^^M{\expandafter\toks0\expandafter{\striprel@x #1}%
\edef\next{\the\toks0}\ifx\next\em@rk\let\next=\endgroup\else\ifx\next\empty%
\else\immediate\write\wfile{\the\toks0}\fi\let\next=\writ@line\fi\next\relax}}
\def\striprel@x#1{} \def\em@rk{\hbox{}}
\def\lref{\begingroup\obeylines\lr@f}
\def\lr@f#1#2{\DefWarn#1\gdef#1{\let#1=\UNd@FiNeD\ref#1{#2}}\endgroup\unskip}

\def\addref#1{\immediate\write\rfile{\noexpand\item{}#1}} 
\def\listrefs{\footatend\vfill\supereject\immediate\closeout\rfile\writestoppt
\baselineskip=\footskip\centerline{{\bf References}}\bigskip{\parindent=20pt%
\frenchspacing\escapechar=` \input \jobname.refs\vfill\eject}\nonfrenchspacing}
\def\startrefs#1{\immediate\openout\rfile=\jobname.refs\refno=#1}
\def\xref{\expandafter\xr@f}\def\xr@f[#1]{#1}
\def\refs#1{\count255=1[\r@fs #1{\hbox{}}]}
\def\r@fs#1{\ifx\UNd@FiNeD#1\message{reflabel \string#1 is undefined.}%
\nref#1{need to supply reference \string#1.}\fi%
\vphantom{\hphantom{#1}}{\let\hyperref=\relax\xdef\next{#1}}%
\ifx\next\em@rk\def\next{}%
\else\ifx\next#1\ifodd\count255\relax\xref#1\count255=0\fi%
\else#1\count255=1\fi\let\next=\r@fs\fi\next}
%

%
\newwrite\ffile\global\newcount\figno \global\figno=1
\def\fig{fig.~\hyperref{}{figure}{\the\figno}{\the\figno}\nfig}
\def\nfig#1{\DefWarn#1%
\xdef#1{fig.~\noexpand\hyperref{}{figure}{\the\figno}{\the\figno}}%
\writedef{#1\leftbracket fig.\noexpand~\xfig#1}%
\ifnum\figno=1\immediate\openout\ffile=\jobname.figs\fi\chardef\wfile=\ffile%
{\let\hyperref=\relax
\immediate\write\ffile{\noexpand\medskip\noexpand\item{Fig.\ %
\noexpand\hyperdef\noexpand\hypernoname{figure}{\the\figno}{\the\figno}. }
\reflabeL{#1\hskip.55in}\pctsign}}\global\advance\figno by1\findarg}
\def\listfigs{\vfill\eject\immediate\closeout\ffile{\parindent40pt
\baselineskip14pt\centerline{{\bf Figure Captions}}\nobreak\medskip
\escapechar=` \input \jobname.figs\vfill\eject}}
\def\xfig{\expandafter\xf@g}\def\xf@g fig.\penalty\@M\ {}
\def\figs#1{figs.~\f@gs #1{\hbox{}}}
\def\f@gs#1{{\let\hyperref=\relax\xdef\next{#1}}\ifx\next\em@rk\def\next{}\else
\ifx\next#1\xfig #1\else#1\fi\let\next=\f@gs\fi\next}
\def\figin{\epsfcheck\figin}\def\figins{\epsfcheck\figins}
\def\epsfcheck{\ifx\epsfbox\UNd@FiNeD
\message{(NO epsf.tex, FIGURES WILL BE IGNORED)}
\gdef\figin##1{\vskip2in}\gdef\figins##1{\hskip.5in}
\else\message{(FIGURES WILL BE INCLUDED)}%
\gdef\figin##1{##1}\gdef\figins##1{##1}\fi}
\def\DefWarn#1{}
\def\figinsert{\goodbreak\midinsert}
\def\ifig#1#2#3{\DefWarn#1\xdef#1{fig.~\noexpand\hyperref{}{figure}%
{\the\figno}{\the\figno}}\writedef{#1\leftbracket fig.\noexpand~\xfig#1}%
\figinsert\figin{\centerline{#3}}\medskip\centerline{\vbox{\baselineskip12pt
\advance\hsize by -1truein\noindent\wrlabeL{#1=#1}\footnotefont%
{\bf Fig.~\hyperdef\hypernoname{figure}{\the\figno}{\the\figno}:} #2}}
\bigskip\endinsert\global\advance\figno by1}
\newwrite\lfile
{\escapechar-1\xdef\pctsign{\string\%}\xdef\leftbracket{\string\{}
\xdef\rightbracket{\string\}}\xdef\numbersign{\string\#}}
\def\writedefs{\immediate\openout\lfile=\jobname.defs \def\writedef##1{%
{\let\hyperref=\relax\let\hyperdef=\relax\let\hypernoname=\relax
 \immediate\write\lfile{\string\def\string##1\rightbracket}}}}%
\def\writestop{\def\writestoppt{\immediate\write\lfile{\string\pageno
 \the\pageno\string\startrefs\leftbracket\the\refno\rightbracket
 \string\def\string\secsym\leftbracket\secsym\rightbracket
 \string\secno\the\secno\string\meqno\the\meqno}\immediate\closeout\lfile}}
\def\writestoppt{}\def\writedef#1{}
\def\seclab#1{\DefWarn#1%
\xdef #1{\noexpand\hyperref{}{section}{\the\secno}{\the\secno}}%
\writedef{#1\leftbracket#1}\wrlabeL{#1=#1}}
\def\subseclab#1{\DefWarn#1%
\xdef #1{\noexpand\hyperref{}{subsection}{\secn@m.\the\subsecno}%
{\secn@m.\the\subsecno}}\writedef{#1\leftbracket#1}\wrlabeL{#1=#1}}
\def\applab#1{\DefWarn#1%
\xdef #1{\noexpand\hyperref{}{appendix}{\secn@m}{\secn@m}}%
\writedef{#1\leftbracket#1}\wrlabeL{#1=#1}}
\newwrite\tfile \def\writetoca#1{}
\def\leaderfill{\leaders\hbox to 1em{\hss.\hss}\hfill}
\def\writetoc{\immediate\openout\tfile=\jobname.toc
   \def\writetoca##1{{\edef\next{\write\tfile{\noindent ##1
   \string\leaderfill {\string\hyperref{}{page}{\noexpand\number\pageno}%
                       {\noexpand\number\pageno}} \par}}\next}}}
\newread\ch@ckfile
\def\listtoc{\immediate\closeout\tfile\immediate\openin\ch@ckfile=\jobname.toc
\ifeof\ch@ckfile\message{no file \jobname.toc, no table of contents this pass}%
\else\closein\ch@ckfile\centerline{\bf Contents}\nobreak\medskip%
{\baselineskip=12pt\footnotefont\parskip=0pt\catcode`\@=11\input\jobname.toc
\catcode`\@=12\bigbreak\bigskip}\fi}
\catcode`\@=12 
%
\edef\tfontsize{\ifx\answ\bigans scaled\magstep3\else scaled\magstep4\fi}
\font\titlerm=cmr10 \tfontsize \font\titlerms=cmr7 \tfontsize
\font\titlermss=cmr5 \tfontsize \font\titlei=cmmi10 \tfontsize
\font\titleis=cmmi7 \tfontsize \font\titleiss=cmmi5 \tfontsize
\font\titlesy=cmsy10 \tfontsize \font\titlesys=cmsy7 \tfontsize
\font\titlesyss=cmsy5 \tfontsize \font\titleit=cmti10 \tfontsize
\skewchar\titlei='177 \skewchar\titleis='177 \skewchar\titleiss='177
\skewchar\titlesy='60 \skewchar\titlesys='60 \skewchar\titlesyss='60
\def\titlefont{\def\rm{\fam0\titlerm}
\textfont0=\titlerm \scriptfont0=\titlerms \scriptscriptfont0=\titlermss
\textfont1=\titlei \scriptfont1=\titleis \scriptscriptfont1=\titleiss
\textfont2=\titlesy \scriptfont2=\titlesys \scriptscriptfont2=\titlesyss
\textfont\itfam=\titleit \def\it{\fam\itfam\titleit}\rm}
 \ifx\answ\bigans\else scaled\magstep1\fi
\ifx\answ\bigans\def\abstractfont{\tenpoint}\else
\font\absit=cmti10 scaled \magstep1
\font\abssl=cmsl10 scaled \magstep1
\font\absrm=cmr10 scaled\magstep1 \font\absrms=cmr7 scaled\magstep1
\font\absrmss=cmr5 scaled\magstep1 \font\absi=cmmi10 scaled\magstep1
\font\absis=cmmi7 scaled\magstep1 \font\absiss=cmmi5 scaled\magstep1
\font\abssy=cmsy10 scaled\magstep1 \font\abssys=cmsy7 scaled\magstep1
\font\abssyss=cmsy5 scaled\magstep1 \font\absbf=cmbx10 scaled\magstep1
\skewchar\absi='177 \skewchar\absis='177 \skewchar\absiss='177
\skewchar\abssy='60 \skewchar\abssys='60 \skewchar\abssyss='60
\def\abstractfont{\def\rm{\fam0\absrm}
\textfont0=\absrm \scriptfont0=\absrms \scriptscriptfont0=\absrmss
\textfont1=\absi \scriptfont1=\absis \scriptscriptfont1=\absiss
\textfont2=\abssy \scriptfont2=\abssys \scriptscriptfont2=\abssyss
\textfont\itfam=\absit \def\it{\fam\itfam\absit}\def\footnotefont{\tenpoint}%
\textfont\slfam=\abssl \def\sl{\fam\slfam\abssl}%
\textfont\bffam=\absbf \def\bf{\fam\bffam\absbf}\rm}\fi
\def\tenpoint{\def\rm{\fam0\tenrm}
\textfont0=\tenrm \scriptfont0=\sevenrm \scriptscriptfont0=\fiverm
\textfont1=\teni  \scriptfont1=\seveni  \scriptscriptfont1=\fivei
\textfont2=\tensy \scriptfont2=\sevensy \scriptscriptfont2=\fivesy
\textfont\itfam=\tenit \def\it{\fam\itfam\tenit}\def\footnotefont{\ninepoint}%
\textfont\bffam=\tenbf \def\bf{\fam\bffam\tenbf}\def\sl{\fam\slfam\tensl}\rm}
\font\ninerm=cmr9 \font\sixrm=cmr6 \font\ninei=cmmi9 \font\sixi=cmmi6
\font\ninesy=cmsy9 \font\sixsy=cmsy6 \font\ninebf=cmbx9
\font\nineit=cmti9 \font\ninesl=cmsl9 \skewchar\ninei='177
\skewchar\sixi='177 \skewchar\ninesy='60 \skewchar\sixsy='60
\def\ninepoint{\def\rm{\fam0\ninerm}
\textfont0=\ninerm \scriptfont0=\sixrm \scriptscriptfont0=\fiverm
\textfont1=\ninei \scriptfont1=\sixi \scriptscriptfont1=\fivei
\textfont2=\ninesy \scriptfont2=\sixsy \scriptscriptfont2=\fivesy
\textfont\itfam=\ninei \def\it{\fam\itfam\nineit}\def\sl{\fam\slfam\ninesl}%
\textfont\bffam=\ninebf \def\bf{\fam\bffam\ninebf}\rm}
%
%

\hyphenation{anom-aly anom-alies coun-ter-term coun-ter-terms}
\def\inv{^{\raise.15ex\hbox{${\scriptscriptstyle -}$}\kern-.05em 1}}

\def\Dsl{\,\raise.15ex\hbox{/}\mkern-13.5mu D} 
\def\dsl{\raise.15ex\hbox{/}\kern-.57em\partial}

\def\lspace{\ifx\answ\bigans{}\else\qquad\fi}
\def\lbspace{\ifx\answ\bigans{}\else\hskip-.2in\fi} 
\def\boxeqn#1{\vcenter{\vbox{\hrule\hbox{\vrule\kern3pt\vbox{\kern3pt
	\hbox{${\displaystyle #1}$}\kern3pt}\kern3pt\vrule}\hrule}}}
\def\mbox#1#2{\vcenter{\hrule \hbox{\vrule height#2in
		\kern#1in \vrule} \hrule}}  
%

\def\darr#1{\raise1.5ex\hbox{$\leftrightarrow$}\mkern-16.5mu #1}

\def\roughly#1{\raise.3ex\hbox{$#1$\kern-.75em\lower1ex\hbox{$\sim$}}}

\def\bb{
\font\tenmsb=msbm10
\font\sevenmsb=msbm7
\font\fivemsb=msbm5
\textfont1=\tenmsb
\scriptfont1=\sevenmsb
\scriptscriptfont1=\fivemsb
}

\input amssym

\input epsf

\def\IZ{\relax\ifmmode\mathchoice
{\hbox{\cmss Z\kern-.4em Z}}{\hbox{\cmss Z\kern-.4em Z}} {\lower.9pt\hbox{\cmsss Z\kern-.4em Z}}
{\lower1.2pt\hbox{\cmsss Z\kern-.4em Z}}\else{\cmss Z\kern-.4em Z}\fi}

\newif\ifdraft\draftfalse
\newif\ifinter\interfalse
\ifdraft\draftmode\else\interfalse\fi
\def\journal#1&#2(#3){\unskip, \sl #1\ \bf #2 \rm(19#3) }
\def\andjournal#1&#2(#3){\sl #1~\bf #2 \rm (19#3) }

\def\ie{{\it i.e.}}

\def\frac#1#2{{#1\over#2}}

\def\inbar{\,\vrule height1.5ex width.4pt depth0pt}
\def\IC{\relax\hbox{$\inbar\kern-.3em{\rm C}$}}
\def\IR{\relax{\rm I\kern-.18em R}}
\def\IP{\relax{\rm I\kern-.18em P}}
\def\Z{{\bf Z}}

%
%


%
\catcode`\@=11
\def\slash#1{\mathord{\mathpalette\c@ncel{#1}}}
\overfullrule=0pt

\def\underrel#1\over#2{\mathrel{\mathop{\kern\z@#1}\limits_{#2}}}

\catcode`\@=12


%

\def \cosh{{\rm cosh}}


\def\[{[}
\def\]{]}

\def\comment#1{ }

%
\def\draftnote#1{\ifdraft{\baselineskip2ex
                 \vbox{\kern1em\hrule\hbox{\vrule\kern1em\vbox{\kern1ex
                 \noindent \underbar{NOTE}: #1
             \vskip1ex}\kern1em\vrule}\hrule}}\fi}
\def\internote#1{\ifinter{\baselineskip2ex
                 \vbox{\kern1em\hrule\hbox{\vrule\kern1em\vbox{\kern1ex
                 \noindent \underbar{Internal Note}: #1
             \vskip1ex}\kern1em\vrule}\hrule}}\fi}

%
%



%
%
%
%

%

\def\inv{^{-1}}


\def\cN{{\cal N}}
\def\cM{{\cal M}}
\def\cZ{{\cal Z}}
\def\1{{\ds 1}}
\def\R{\hbox{$\bb R$}}
\def\C{\hbox{$\bb C$}}

\def\Z{\hbox{$\bb Z$}}

\def\S{\hbox{$\bb S$}}

\newfam\frakfam
\font\teneufm=eufm10
\font\seveneufm=eufm7
\font\fiveeufm=eufm5
\textfont\frakfam=\teneufm
\scriptfont\frakfam=\seveneufm
\scriptscriptfont\frakfam=\fiveeufm

\lref\NiarchosAH{
  V.~Niarchos,
  ``Seiberg dualities and the 3d/4d connection,''
JHEP {\bf 1207}, 075 (2012).
[arXiv:1205.2086 [hep-th]].
}

\lref\AharonyGP{
  O.~Aharony,
  ``IR duality in d = 3 N=2 supersymmetric USp(2N(c)) and U(N(c)) gauge theories,''
Phys.\ Lett.\ B {\bf 404}, 71 (1997).
[hep-th/9703215].
}

\lref\AffleckAS{
  I.~Affleck, J.~A.~Harvey and E.~Witten,
  ``Instantons and (Super)Symmetry Breaking in (2+1)-Dimensions,''
Nucl.\ Phys.\ B {\bf 206}, 413 (1982)..
}

\lref\IntriligatorID{
  K.~A.~Intriligator and N.~Seiberg,
  ``Duality, monopoles, dyons, confinement and oblique confinement in supersymmetric SO(N(c)) gauge theories,''
Nucl.\ Phys.\ B {\bf 444}, 125 (1995).
[hep-th/9503179].
}

\lref\PasquettiFJ{
  S.~Pasquetti,
  ``Factorisation of N = 2 Theories on the Squashed 3-Sphere,''
JHEP {\bf 1204}, 120 (2012).
[arXiv:1111.6905 [hep-th]].
}

\lref\BeemMB{
  C.~Beem, T.~Dimofte and S.~Pasquetti,
  ``Holomorphic Blocks in Three Dimensions,''
[arXiv:1211.1986 [hep-th]].
}

\lref\SeibergPQ{
  N.~Seiberg,
  ``Electric - magnetic duality in supersymmetric nonAbelian gauge theories,''
Nucl.\ Phys.\ B {\bf 435}, 129 (1995).
[hep-th/9411149].
}

\lref\AharonyBX{
  O.~Aharony, A.~Hanany, K.~A.~Intriligator, N.~Seiberg and M.~J.~Strassler,
  ``Aspects of N=2 supersymmetric gauge theories in three-dimensions,''
Nucl.\ Phys.\ B {\bf 499}, 67 (1997).
[hep-th/9703110].
}

\lref\IntriligatorNE{
  K.~A.~Intriligator and P.~Pouliot,
  ``Exact superpotentials, quantum vacua and duality in supersymmetric SP(N(c)) gauge theories,''
Phys.\ Lett.\ B {\bf 353}, 471 (1995).
[hep-th/9505006].
}

\lref\KarchUX{
  A.~Karch,
  ``Seiberg duality in three-dimensions,''
Phys.\ Lett.\ B {\bf 405}, 79 (1997).
[hep-th/9703172].
}

\lref\SafdiRE{
  B.~R.~Safdi, I.~R.~Klebanov and J.~Lee,
  ``A Crack in the Conformal Window,''
[arXiv:1212.4502 [hep-th]].
}

\lref\AharonyTH{
  O.~Aharony, M.~Berkooz, S.~Kachru, N.~Seiberg and E.~Silverstein,
  ``Matrix description of interacting theories in six-dimensions,''
Adv.\ Theor.\ Math.\ Phys.\  {\bf 1}, 148 (1998).
[hep-th/9707079].
}

\lref\SchweigertTG{
  C.~Schweigert,
  ``On moduli spaces of flat connections with nonsimply connected structure group,''
Nucl.\ Phys.\ B {\bf 492}, 743 (1997).
[hep-th/9611092].
}

\lref\WittenYU{
  E.~Witten,
  ``On the conformal field theory of the Higgs branch,''
JHEP {\bf 9707}, 003 (1997).
[hep-th/9707093].
}

\lref\GiveonZN{
  A.~Giveon and D.~Kutasov,
  ``Seiberg Duality in Chern-Simons Theory,''
Nucl.\ Phys.\ B {\bf 812}, 1 (2009).
[arXiv:0808.0360 [hep-th]].
}

\lref\GaiottoBE{
  D.~Gaiotto, G.~W.~Moore and A.~Neitzke,
  ``Framed BPS States,''
[arXiv:1006.0146 [hep-th]].
}

\lref\SethiPA{
  S.~Sethi and M.~Stern,
  ``D-brane bound states redux,''
Commun.\ Math.\ Phys.\  {\bf 194}, 675 (1998).
[hep-th/9705046].
}

\lref\AldayRS{
  L.~F.~Alday, M.~Bullimore and M.~Fluder,
  ``On S-duality of the Superconformal Index on Lens Spaces and 2d TQFT,''
JHEP {\bf 1305}, 122 (2013).
[arXiv:1301.7486 [hep-th]].
}

\lref\RazamatJXA{
  S.~S.~Razamat and M.~Yamazaki,
  ``S-duality and the N=2 Lens Space Index,''
[arXiv:1306.1543 [hep-th]].
}

\lref\HoriDK{
  K.~Hori and D.~Tong,
  ``Aspects of Non-Abelian Gauge Dynamics in Two-Dimensional N=(2,2) Theories,''
JHEP {\bf 0705}, 079 (2007).
[hep-th/0609032].
}

\lref\NiarchosAH{
  V.~Niarchos,
  ``Seiberg dualities and the 3d/4d connection,''
JHEP {\bf 1207}, 075 (2012).
[arXiv:1205.2086 [hep-th]].
}

\lref\almost{
  A.~Borel, R.~Friedman, J.~W.~Morgan,
  ``Almost commuting elements in compact Lie groups,''
arXiv:math/9907007.
}

\lref\KapustinJM{
  A.~Kapustin and B.~Willett,
  ``Generalized Superconformal Index for Three Dimensional Field Theories,''
[arXiv:1106.2484 [hep-th]].
}

\lref\AharonyGP{
  O.~Aharony,
  ``IR duality in d = 3 N=2 supersymmetric USp(2N(c)) and U(N(c)) gauge theories,''
Phys.\ Lett.\ B {\bf 404}, 71 (1997).
[hep-th/9703215].
}

\lref\FestucciaWS{
  G.~Festuccia and N.~Seiberg,
  ``Rigid Supersymmetric Theories in Curved Superspace,''
JHEP {\bf 1106}, 114 (2011).
[arXiv:1105.0689 [hep-th]].
}

\lref\RomelsbergerEG{
  C.~Romelsberger,
  ``Counting chiral primaries in N = 1, d=4 superconformal field theories,''
Nucl.\ Phys.\ B {\bf 747}, 329 (2006).
[hep-th/0510060].
}

\lref\KapustinKZ{
  A.~Kapustin, B.~Willett and I.~Yaakov,
  ``Exact Results for Wilson Loops in Superconformal Chern-Simons Theories with Matter,''
JHEP {\bf 1003}, 089 (2010).
[arXiv:0909.4559 [hep-th]].
}

\lref\DolanQI{
  F.~A.~Dolan and H.~Osborn,
  ``Applications of the Superconformal Index for Protected Operators and q-Hypergeometric Identities to N=1 Dual Theories,''
Nucl.\ Phys.\ B {\bf 818}, 137 (2009).
[arXiv:0801.4947 [hep-th]].
}

\lref\GaddeIA{
  A.~Gadde and W.~Yan,
  ``Reducing the 4d Index to the $S^3$ Partition Function,''
JHEP {\bf 1212}, 003 (2012).
[arXiv:1104.2592 [hep-th]].
}

\lref\DolanRP{
  F.~A.~H.~Dolan, V.~P.~Spiridonov and G.~S.~Vartanov,
  ``From 4d superconformal indices to 3d partition functions,''
Phys.\ Lett.\ B {\bf 704}, 234 (2011).
[arXiv:1104.1787 [hep-th]].
}

\lref\ImamuraUW{
  Y.~Imamura,
 ``Relation between the 4d superconformal index and the $S^3$ partition function,''
JHEP {\bf 1109}, 133 (2011).
[arXiv:1104.4482 [hep-th]].
}

\lref\HamaEA{
  N.~Hama, K.~Hosomichi and S.~Lee,
  ``SUSY Gauge Theories on Squashed Three-Spheres,''
JHEP {\bf 1105}, 014 (2011).
[arXiv:1102.4716 [hep-th]].
}

\lref\GaddeEN{
  A.~Gadde, L.~Rastelli, S.~S.~Razamat and W.~Yan,
  ``On the Superconformal Index of N=1 IR Fixed Points: A Holographic Check,''
JHEP {\bf 1103}, 041 (2011).
[arXiv:1011.5278 [hep-th]].
}

\lref\EagerHX{
  R.~Eager, J.~Schmude and Y.~Tachikawa,
  ``Superconformal Indices, Sasaki-Einstein Manifolds, and Cyclic Homologies,''
[arXiv:1207.0573 [hep-th]].
}

\lref\AffleckAS{
  I.~Affleck, J.~A.~Harvey and E.~Witten,
  ``Instantons and (Super)Symmetry Breaking in (2+1)-Dimensions,''
Nucl.\ Phys.\ B {\bf 206}, 413 (1982)..
}

\lref\SeibergPQ{
  N.~Seiberg,
  ``Electric - magnetic duality in supersymmetric nonAbelian gauge theories,''
Nucl.\ Phys.\ B {\bf 435}, 129 (1995).
[hep-th/9411149].
}

\lref\AlvarezGaumeNF{
  L.~Alvarez-Gaume, S.~Della Pietra and G.~W.~Moore,
  ``Anomalies and Odd Dimensions,''
Annals Phys.\  {\bf 163}, 288 (1985)..
}

\lref\SeibergRSG{
  N.~Seiberg and E.~Witten,
  ``Gapped Boundary Phases of Topological Insulators via Weak Coupling,''
[arXiv:1602.04251 [cond-mat.str-el]].
}

\lref\debult{
  F.~van~de~Bult,
  ``Hyperbolic Hypergeometric Functions,''
University of Amsterdam Ph.D. thesis
}

\lref\Shamirthesis{
  I.~Shamir,
  ``Aspects of three dimensional Seiberg duality,''
  M. Sc. thesis submitted to the Weizmann Institute of Science, April 2010.
  }

\lref\slthreeZ{
  J.~Felder, A.~Varchenko,
  ``The elliptic gamma function and $SL(3,Z) \times Z^3$,'' $\;\;$
[arXiv:math/0001184].
}

\lref\BeniniNC{
  F.~Benini, T.~Nishioka and M.~Yamazaki,
  ``4d Index to 3d Index and 2d TQFT,''
Phys.\ Rev.\ D {\bf 86}, 065015 (2012).
[arXiv:1109.0283 [hep-th]].
}

\lref\GaddeDDA{
  A.~Gadde and S.~Gukov,
  ``2d Index and Surface operators,''
JHEP {\bf 1403}, 080 (2014).
[arXiv:1305.0266 [hep-th]].
}

\lref\GaiottoWE{
  D.~Gaiotto,
  ``N=2 dualities,''
  JHEP {\bf 1208}, 034 (2012).
  [arXiv:0904.2715 [hep-th]].
}

\lref\WittenZH{
  E.~Witten,
  ``Some comments on string dynamics,''
In *Los Angeles 1995, Future perspectives in string theory* 501-523.
[hep-th/9507121].
}

\lref\SpiridonovZA{
  V.~P.~Spiridonov and G.~S.~Vartanov,
  ``Elliptic Hypergeometry of Supersymmetric Dualities,''
Commun.\ Math.\ Phys.\  {\bf 304}, 797 (2011).
[arXiv:0910.5944 [hep-th]].
}

\lref\BeniniMF{
  F.~Benini, C.~Closset and S.~Cremonesi,
  ``Comments on 3d Seiberg-like dualities,''
JHEP {\bf 1110}, 075 (2011).
[arXiv:1108.5373 [hep-th]].
}

\lref\ClossetVP{
  C.~Closset, T.~T.~Dumitrescu, G.~Festuccia, Z.~Komargodski and N.~Seiberg,
  ``Comments on Chern-Simons Contact Terms in Three Dimensions,''
JHEP {\bf 1209}, 091 (2012).
[arXiv:1206.5218 [hep-th]].
}

\lref\SpiridonovHF{
  V.~P.~Spiridonov and G.~S.~Vartanov,
  ``Elliptic hypergeometry of supersymmetric dualities II. Orthogonal groups, knots, and vortices,''
[arXiv:1107.5788 [hep-th]].
}

\lref\SpiridonovWW{
  V.~P.~Spiridonov and G.~S.~Vartanov,
  ``Elliptic hypergeometric integrals and 't Hooft anomaly matching conditions,''
JHEP {\bf 1206}, 016 (2012).
[arXiv:1203.5677 [hep-th]].
}

\lref\DimoftePY{
  T.~Dimofte, D.~Gaiotto and S.~Gukov,
  ``3-Manifolds and 3d Indices,''
[arXiv:1112.5179 [hep-th]].
}

\lref\KimWB{
  S.~Kim,
  ``The Complete superconformal index for N=6 Chern-Simons theory,''
Nucl.\ Phys.\ B {\bf 821}, 241 (2009), [Erratum-ibid.\ B {\bf 864}, 884 (2012)].
[arXiv:0903.4172 [hep-th]].
}

\lref\WillettGP{
  B.~Willett and I.~Yaakov,
  ``N=2 Dualities and Z Extremization in Three Dimensions,''
[arXiv:1104.0487 [hep-th]].
}

\lref\ImamuraSU{
  Y.~Imamura and S.~Yokoyama,
  ``Index for three dimensional superconformal field theories with general R-charge assignments,''
JHEP {\bf 1104}, 007 (2011).
[arXiv:1101.0557 [hep-th]].
}

\lref\FreedYA{
  D.~S.~Freed, G.~W.~Moore and G.~Segal,
  ``The Uncertainty of Fluxes,''
Commun.\ Math.\ Phys.\  {\bf 271}, 247 (2007).
[hep-th/0605198].
}

\lref\HwangQT{
  C.~Hwang, H.~Kim, K.~-J.~Park and J.~Park,
  ``Index computation for 3d Chern-Simons matter theory: test of Seiberg-like duality,''
JHEP {\bf 1109}, 037 (2011).
[arXiv:1107.4942 [hep-th]].
}

\lref\ParkWTA{
  J.~Park and K.~J.~Park,
  ``Seiberg-like Dualities for 3d N=2 Theories with SU(N) gauge group,''
[arXiv:1305.6280 [hep-th]].
}

\lref\JafferisNS{
  D.~Jafferis and X.~Yin,
  ``A Duality Appetizer,''
[arXiv:1103.5700 [hep-th]].
}

\lref\GreenDA{
  D.~Green, Z.~Komargodski, N.~Seiberg, Y.~Tachikawa and B.~Wecht,
  ``Exactly Marginal Deformations and Global Symmetries,''
JHEP {\bf 1006}, 106 (2010).
[arXiv:1005.3546 [hep-th]].
}

\lref\GaiottoXA{
  D.~Gaiotto, L.~Rastelli and S.~S.~Razamat,
  ``Bootstrapping the superconformal index with surface defects,''
[arXiv:1207.3577 [hep-th]].
}

\lref\TroostUD{
  J.~Troost,
 ``The non-compact elliptic genus: mock or modular,''
JHEP {\bf 1006}, 104 (2010).
[arXiv:1004.3649 [hep-th]].
}

\lref\BhattacharyaZY{
  J.~Bhattacharya, S.~Bhattacharyya, S.~Minwalla and S.~Raju,
  ``Indices for Superconformal Field Theories in 3,5 and 6 Dimensions,''
JHEP {\bf 0802}, 064 (2008).
[arXiv:0801.1435 [hep-th]].
}

\lref\IntriligatorID{
  K.~A.~Intriligator and N.~Seiberg,
  ``Duality, monopoles, dyons, confinement and oblique confinement in supersymmetric SO(N(c)) gauge theories,''
Nucl.\ Phys.\ B {\bf 444}, 125 (1995).
[hep-th/9503179].
}

\lref\SeibergNZ{
  N.~Seiberg and E.~Witten,
  ``Gauge dynamics and compactification to three-dimensions,''
In *Saclay 1996, The mathematical beauty of physics* 333-366.
[hep-th/9607163].
}

\lref\KinneyEJ{
  J.~Kinney, J.~M.~Maldacena, S.~Minwalla and S.~Raju,
  ``An Index for 4 dimensional super conformal theories,''
  Commun.\ Math.\ Phys.\  {\bf 275}, 209 (2007).
  [hep-th/0510251].
}

\lref\NakayamaUR{
  Y.~Nakayama,
  ``Index for supergravity on AdS(5) x T**1,1 and conifold gauge theory,''
Nucl.\ Phys.\ B {\bf 755}, 295 (2006).
[hep-th/0602284].
}

\lref\GaddeKB{
  A.~Gadde, E.~Pomoni, L.~Rastelli and S.~S.~Razamat,
  ``S-duality and 2d Topological QFT,''
JHEP {\bf 1003}, 032 (2010).
[arXiv:0910.2225 [hep-th]].
}

\lref\GaddeTE{
  A.~Gadde, L.~Rastelli, S.~S.~Razamat and W.~Yan,
  ``The Superconformal Index of the $E_6$ SCFT,''
JHEP {\bf 1008}, 107 (2010).
[arXiv:1003.4244 [hep-th]].
}

\lref\AharonyCI{
  O.~Aharony and I.~Shamir,
  ``On $O(N_c)$ d=3 N=2 supersymmetric QCD Theories,''
JHEP {\bf 1112}, 043 (2011).
[arXiv:1109.5081 [hep-th]].
}

\lref\ClossetVG{
  C.~Closset, T.~T.~Dumitrescu, G.~Festuccia, Z.~Komargodski and N.~Seiberg,
  ``Contact Terms, Unitarity, and F-Maximization in Three-Dimensional Superconformal Theories,''
JHEP {\bf 1210}, 053 (2012).
[arXiv:1205.4142 [hep-th]].
}

\lref\ChenPHA{
  H.~Y.~Chen, H.~Y.~Chen and J.~K.~Ho,
  ``Connecting Mirror Symmetry in 3d and 2d via Localization,''
[arXiv:1312.2361 [hep-th]].
}

\lref\GiveonSR{
  A.~Giveon and D.~Kutasov,
  ``Brane dynamics and gauge theory,''
Rev.\ Mod.\ Phys.\  {\bf 71}, 983 (1999).
[hep-th/9802067].
}

\lref\SpiridonovQV{
  V.~P.~Spiridonov and G.~S.~Vartanov,
  ``Superconformal indices of ${\cal N}=4$ SYM field theories,''
Lett.\ Math.\ Phys.\  {\bf 100}, 97 (2012).
[arXiv:1005.4196 [hep-th]].
}
\lref\GaddeUV{
  A.~Gadde, L.~Rastelli, S.~S.~Razamat and W.~Yan,
  ``Gauge Theories and Macdonald Polynomials,''
Commun.\ Math.\ Phys.\  {\bf 319}, 147 (2013).
[arXiv:1110.3740 [hep-th]].
}
\lref\KapustinGH{
  A.~Kapustin,
  ``Seiberg-like duality in three dimensions for orthogonal gauge groups,''
[arXiv:1104.0466 [hep-th]].
}

\lref\orthogpaper{O. Aharony, S. S. Razamat, N.~Seiberg and B.~Willett,
``3d dualities from 4d dualities for orthogonal groups,''
[arXiv:1307.0511 [hep-th]].
}

\lref\KapustinHA{
  A.~Kapustin and M.~J.~Strassler,
  ``On mirror symmetry in three-dimensional Abelian gauge theories,''
JHEP {\bf 9904}, 021 (1999).
[hep-th/9902033].
}

\lref\IntriligatorEX{
  K.~A.~Intriligator and N.~Seiberg,
  ``Mirror symmetry in three-dimensional gauge theories,''
Phys.\ Lett.\ B {\bf 387}, 513 (1996).
[hep-th/9607207].
}

\lref\deBoerMP{
  J.~de Boer, K.~Hori, H.~Ooguri and Y.~Oz,
  ``Mirror symmetry in three-dimensional gauge theories, quivers and D-branes,''
Nucl.\ Phys.\ B {\bf 493}, 101 (1997).
[hep-th/9611063].
}

\lref\readinglines{
  O.~Aharony, N.~Seiberg and Y.~Tachikawa,
  ``Reading between the lines of four-dimensional gauge theories,''
[arXiv:1305.0318 [hep-th]].
}

\lref\AharonyKMA{
  O.~Aharony, S.~S.~Razamat, N.~Seiberg and B.~Willett,
  ``3$d$ dualities from 4$d$ dualities for orthogonal groups,''
JHEP {\bf 1308}, 099 (2013).
[arXiv:1307.0511, arXiv:1307.0511 [hep-th]].
}

\lref\RazamatOPA{
  S.~S.~Razamat and B.~Willett,
  ``Global Properties of Supersymmetric Theories and the Lens Space,''
[arXiv:1307.4381].
}

\lref\WittenNV{
  E.~Witten,
  ``Supersymmetric index in four-dimensional gauge theories,''
Adv.\ Theor.\ Math.\ Phys.\  {\bf 5}, 841 (2002).
[hep-th/0006010].
}

\lref\BeniniNC{
  F.~Benini, T.~Nishioka and M.~Yamazaki,
  ``4d Index to 3d Index and 2d TQFT,''
Phys.\ Rev.\ D {\bf 86}, 065015 (2012).
[arXiv:1109.0283 [hep-th]].
}

\lref\GaddeUV{
  A.~Gadde, L.~Rastelli, S.~S.~Razamat and W.~Yan,
  ``Gauge Theories and Macdonald Polynomials,''
Commun.\ Math.\ Phys.\  {\bf 319}, 147 (2013).
[arXiv:1110.3740 [hep-th]].
}

\lref\GaddeIK{
  A.~Gadde, L.~Rastelli, S.~S.~Razamat and W.~Yan,
  ``The 4d Superconformal Index from q-deformed 2d Yang-Mills,''
Phys.\ Rev.\ Lett.\  {\bf 106}, 241602 (2011).
[arXiv:1104.3850 [hep-th]].
}

\lref\GaiottoUQ{
  D.~Gaiotto and S.~S.~Razamat,
  ``Exceptional Indices,''
JHEP {\bf 1205}, 145 (2012).
[arXiv:1203.5517 [hep-th]].
}

\lref\JafferisUN{
  D.~L.~Jafferis,
  ``The Exact Superconformal R-Symmetry Extremizes Z,''
JHEP {\bf 1205}, 159 (2012).
[arXiv:1012.3210 [hep-th]].
}

\lref\RazamatUV{
  S.~S.~Razamat,
  ``On a modular property of N=2 superconformal theories in four dimensions,''
JHEP {\bf 1210}, 191 (2012).
[arXiv:1208.5056 [hep-th]].
}

\lref\noumi{
  Y.~Komori, M.~Noumi, J.~Shiraishi,
  ``Kernel Functions for Difference Operators of Ruijsenaars Type and Their Applications,''
SIGMA 5 (2009), 054.
[arXiv:0812.0279 [math.QA]].
}

\lref\RazamatJXA{
  S.~S.~Razamat and M.~Yamazaki,
  ``S-duality and the N=2 Lens Space Index,''
[arXiv:1306.1543 [hep-th]].
}

\lref\RazamatOPA{
  S.~S.~Razamat and B.~Willett,
  ``Global Properties of Supersymmetric Theories and the Lens Space,''
[arXiv:1307.4381 [hep-th]].
}

\lref\GaddeTE{
  A.~Gadde, L.~Rastelli, S.~S.~Razamat and W.~Yan,
  ``The Superconformal Index of the $E_6$ SCFT,''
JHEP {\bf 1008}, 107 (2010).
[arXiv:1003.4244 [hep-th]].
}

\lref\deBult{
  F.~J.~van~de~Bult,
  ``An elliptic hypergeometric integral with $W(F_4)$ symmetry,''
The Ramanujan Journal, Volume 25, Issue 1 (2011)
[arXiv:0909.4793[math.CA]].
}

\lref\GaddeKB{
  A.~Gadde, E.~Pomoni, L.~Rastelli and S.~S.~Razamat,
  ``S-duality and 2d Topological QFT,''
JHEP {\bf 1003}, 032 (2010).
[arXiv:0910.2225 [hep-th]].
}

\lref\ArgyresCN{
  P.~C.~Argyres and N.~Seiberg,
  ``S-duality in N=2 supersymmetric gauge theories,''
JHEP {\bf 0712}, 088 (2007).
[arXiv:0711.0054 [hep-th]].
}

\lref\SpirWarnaar{
  V.~P.~Spiridonov and S.~O.~Warnaar,
  ``Inversions of integral operators and elliptic beta integrals on root systems,''
Adv. Math. 207 (2006), 91-132
[arXiv:math/0411044].
}

\lref\SethiPA{
  S.~Sethi and M.~Stern,
 ``D-brane bound states redux,''
Commun.\ Math.\ Phys.\  {\bf 194}, 675 (1998).
[hep-th/9705046].
}

\lref\GerchkovitzGTA{
  E.~Gerchkovitz, J.~Gomis and Z.~Komargodski,
 ``Sphere Partition Functions and the Zamolodchikov Metric,''
[arXiv:1405.7271 [hep-th]].
}

\lref\GaiottoHG{
  D.~Gaiotto, G.~W.~Moore and A.~Neitzke,
  ``Wall-crossing, Hitchin Systems, and the WKB Approximation,''
[arXiv:0907.3987 [hep-th]].
}

\lref\RuijsenaarsVQ{
  S.~N.~M.~Ruijsenaars and H.~Schneider,
  ``A New Class Of Integrable Systems And Its Relation To Solitons,''
Annals Phys.\  {\bf 170}, 370 (1986).
}

\lref\GaiottoAK{
  D.~Gaiotto and E.~Witten,
  ``S-Duality of Boundary Conditions In N=4 Super Yang-Mills Theory,''
Adv.\ Theor.\ Math.\ Phys.\  {\bf 13}, 721 (2009).
[arXiv:0807.3720 [hep-th]].
}

\lref\RuijsenaarsPP{
  S.~N.~M.~Ruijsenaars,
  ``Complete Integrability Of Relativistic Calogero-moser Systems And Elliptic Function Identities,''
Commun.\ Math.\ Phys.\  {\bf 110}, 191 (1987).
}

\lref\HallnasNB{
  M.~Hallnas and S.~Ruijsenaars,
  ``Kernel functions and Baecklund transformations for relativistic Calogero-Moser and Toda systems,''
J.\ Math.\ Phys.\  {\bf 53}, 123512 (2012).
}

\lref\kernelA{
S.~Ruijsenaars,
  ``Elliptic integrable systems of Calogero-Moser type: Some new results on joint eigenfunctions'', in Proceedings of the 2004 Kyoto Workshop on "Elliptic integrable systems", (M. Noumi, K. Takasaki, Eds.), Rokko Lectures in Math., no. 18, Dept. of Math., Kobe Univ.
}

\lref\ellRSreview{
Y.~Komori and S.~Ruijsenaars,
  ``Elliptic integrable systems of Calogero-Moser type: A survey'', in Proceedings of the 2004 Kyoto Workshop on "Elliptic integrable systems", (M. Noumi, K. Takasaki, Eds.), Rokko Lectures in Math., no. 18, Dept. of Math., Kobe Univ.
}

\lref\langmann{
E.~Langmann,
  ``An explicit solution of the (quantum) elliptic Calogero-Sutherland model'', [arXiv:math-ph/0407050].
}

\lref\TachikawaWI{
  Y.~Tachikawa,
  ``4d partition function on $S^1 \times S^3$ and 2d Yang-Mills with nonzero area,''
PTEP {\bf 2013}, 013B01 (2013).
[arXiv:1207.3497 [hep-th]].
}

\lref\MinahanFG{
  J.~A.~Minahan and D.~Nemeschansky,
  ``An N=2 superconformal fixed point with E(6) global symmetry,''
Nucl.\ Phys.\ B {\bf 482}, 142 (1996).
[hep-th/9608047].
}

\lref\AldayKDA{
  L.~F.~Alday, M.~Bullimore, M.~Fluder and L.~Hollands,
  ``Surface defects, the superconformal index and q-deformed Yang-Mills,''
[arXiv:1303.4460 [hep-th]].
}

\lref\FukudaJR{
  Y.~Fukuda, T.~Kawano and N.~Matsumiya,
  ``5D SYM and 2D q-Deformed YM,''
Nucl.\ Phys.\ B {\bf 869}, 493 (2013).
[arXiv:1210.2855 [hep-th]].
}

\lref\XieHS{
  D.~Xie,
  ``General Argyres-Douglas Theory,''
JHEP {\bf 1301}, 100 (2013).
[arXiv:1204.2270 [hep-th]].
}

\lref\DrukkerSR{
  N.~Drukker, T.~Okuda and F.~Passerini,
  ``Exact results for vortex loop operators in 3d supersymmetric theories,''
[arXiv:1211.3409 [hep-th]].
}

\lref\qinteg{
  M.~Rahman, A.~Verma,
  ``A q-integral representation of Rogers' q-ultraspherical polynomials and some applications,''
Constructive Approximation
1986, Volume 2, Issue 1.
}

\lref\qintegOK{
  A.~Okounkov,
  ``(Shifted) Macdonald Polynomials: q-Integral Representation and Combinatorial Formula,''
Compositio Mathematica
June 1998, Volume 112, Issue 2.
[arXiv:q-alg/9605013].
}

\lref\macNest{
 H.~Awata, S.~Odake, J.~Shiraishi,
  ``Integral Representations of the Macdonald Symmetric Functions,''
Commun. Math. Phys. 179 (1996) 647.
[arXiv:q-alg/9506006].
}

\lref\BeemYN{
  C.~Beem and A.~Gadde,
  ``The superconformal index of N=1 class S fixed points,''
[arXiv:1212.1467 [hep-th]].
}

\lref\GaddeFMA{
  A.~Gadde, K.~Maruyoshi, Y.~Tachikawa and W.~Yan,
  ``New N=1 Dualities,''
JHEP {\bf 1306}, 056 (2013).
[arXiv:1303.0836 [hep-th]].
}

\lref\BeniniNDA{
  F.~Benini, R.~Eager, K.~Hori and Y.~Tachikawa,
  ``Elliptic genera of two-dimensional N=2 gauge theories with rank-one gauge groups,''
Lett.\ Math.\ Phys.\  {\bf 104}, 465 (2014).
[arXiv:1305.0533 [hep-th]].
}

\lref\NekrasovUH{
  N.~A.~Nekrasov and S.~L.~Shatashvili,
  ``Supersymmetric vacua and Bethe ansatz,''
Nucl.\ Phys.\ Proc.\ Suppl.\  {\bf 192-193}, 91 (2009).
[arXiv:0901.4744 [hep-th]].
}

\lref\GorskyTN{
  A.~Gorsky,
  ``Dualities in integrable systems and N=2 SUSY theories,''
J.\ Phys.\ A {\bf 34}, 2389 (2001).
[hep-th/9911037].
}

\lref\IntriligatorLCA{
  K.~Intriligator and N.~Seiberg,
  ``Aspects of 3d N=2 Chern-Simons-Matter Theories,''
JHEP {\bf 1307}, 079 (2013).
[arXiv:1305.1633 [hep-th]].
}

\lref\FockAE{
  V.~Fock, A.~Gorsky, N.~Nekrasov and V.~Rubtsov,
  ``Duality in integrable systems and gauge theories,''
JHEP {\bf 0007}, 028 (2000).
[hep-th/9906235].
}

\lref\CsakiCU{
  C.~Csaki, M.~Schmaltz, W.~Skiba and J.~Terning,
  ``Selfdual N=1 SUSY gauge theories,''
Phys.\ Rev.\ D {\bf 56}, 1228 (1997).
[hep-th/9701191].
}

\lref\BeniniUI{
  F.~Benini and S.~Cremonesi,
  ``Partition functions of $N=(2,2)$ gauge theories on $S^2$ and vortices,''
Commun.\ Math.\ Phys.\  {\bf 334}, no. 3, 1483 (2015).
[arXiv:1206.2356 [hep-th]].
}

\lref\DoroudXW{
  N.~Doroud, J.~Gomis, B.~Le Floch and S.~Lee,
  ``Exact Results in D=2 Supersymmetric Gauge Theories,''
JHEP {\bf 1305}, 093 (2013).
[arXiv:1206.2606 [hep-th]].
}

\lref\GomisWY{
  J.~Gomis and S.~Lee,
  ``Exact Kahler Potential from Gauge Theory and Mirror Symmetry,''
JHEP {\bf 1304}, 019 (2013).
[arXiv:1210.6022 [hep-th]].
}

\lref\AganagicUW{
  M.~Aganagic, K.~Hori, A.~Karch and D.~Tong,
  ``Mirror symmetry in (2+1)-dimensions and (1+1)-dimensions,''
JHEP {\bf 0107}, 022 (2001).
[hep-th/0105075].
}

\lref\BeniniNDA{
  F.~Benini, R.~Eager, K.~Hori and Y.~Tachikawa,
  ``Elliptic genera of two-dimensional N=2 gauge theories with rank-one gauge groups,''
Lett.\ Math.\ Phys.\  {\bf 104}, 465 (2014).
[arXiv:1305.0533 [hep-th]].
}

\lref\HoriKT{
  K.~Hori and C.~Vafa,
  ``Mirror symmetry,''
[hep-th/0002222].
}

\lref\SeibergBD{
  N.~Seiberg,
  ``Five-dimensional SUSY field theories, nontrivial fixed points and string dynamics,''
Phys.\ Lett.\ B {\bf 388}, 753 (1996).
[hep-th/9608111].
}

\lref\BeniniMIA{
  F.~Benini, D.~S.~Park and P.~Zhao,
  ``Cluster Algebras from Dualities of 2d N = (2,~2) Quiver Gauge Theories,''
Commun.\ Math.\ Phys.\  {\bf 340}, 47 (2015).
[arXiv:1406.2699 [hep-th]].
}

\lref\AharonyDHA{
  O.~Aharony, S.~S.~Razamat, N.~Seiberg and B.~Willett,
  ``3d dualities from 4d dualities,''
JHEP {\bf 1307}, 149 (2013).
[arXiv:1305.3924 [hep-th]].
}

\lref\HoriPD{
  K.~Hori,
  ``Duality In Two-Dimensional (2,2) Supersymmetric Non-Abelian Gauge Theories,''
JHEP {\bf 1310}, 121 (2013).
[arXiv:1104.2853 [hep-th]].
}

\lref\DoroudPKA{
  N.~Doroud and J.~Gomis,
  ``Gauge Theory Dynamics and Kahler Potential for Calabi-Yau Complex Moduli,''
[arXiv:1309.2305 [hep-th]].
}

\lref\WittenYC{
  E.~Witten,
  ``Phases of N=2 theories in two-dimensions,''
Nucl.\ Phys.\ B {\bf 403}, 159 (1993).
[hep-th/9301042].
}

\lref\GerchkovitzGTA{
  E.~Gerchkovitz, J.~Gomis and Z.~Komargodski,
  ``Sphere Partition Functions and the Zamolodchikov Metric,''
[arXiv:1405.7271 [hep-th]].
}

\lref\ImamuraSU{
  Y.~Imamura and S.~Yokoyama,
  ``Index for three dimensional superconformal field theories with general R-charge assignments,''
JHEP {\bf 1104}, 007 (2011).
[arXiv:1101.0557 [hep-th]].
}

\lref\DimofteJU{
  T.~Dimofte, D.~Gaiotto and S.~Gukov,
  ``Gauge Theories Labelled by Three-Manifolds,''
Commun.\ Math.\ Phys.\  {\bf 325}, 367 (2014).
[arXiv:1108.4389 [hep-th]].
}

\lref\qformsym{
D.~Gaiotto, A.~Kapustin, N.~Seiberg, B.~Willett,
``Generalized global symmetries,''
{\it to appear}
}

\lref\BeniniXPA{
  F.~Benini, R.~Eager, K.~Hori and Y.~Tachikawa,
  ``Elliptic Genera of 2d $N = 2$ Gauge Theories,''
Commun.\ Math.\ Phys.\  {\bf 333}, no. 3, 1241 (2015).
[arXiv:1308.4896 [hep-th]].
}

\lref\DoreyRB{
  N.~Dorey and D.~Tong,
  ``Mirror symmetry and toric geometry in three-dimensional gauge theories,''
JHEP {\bf 0005}, 018 (2000).
[hep-th/9911094].
}

\lref\DiPietroBCA{
  L.~Di Pietro and Z.~Komargodski,
  ``Cardy Formulae for SUSY Theories in d=4 and d=6,''
[arXiv:1407.6061 [hep-th]].
}

\lref\HoriAX{
  K.~Hori and A.~Kapustin,
  ``Duality of the fermionic 2-D black hole and N=2 liouville theory as mirror symmetry,''
JHEP {\bf 0108}, 045 (2001).
[hep-th/0104202].
}

\lref\GiveonPX{
  A.~Giveon and D.~Kutasov,
  ``Little string theory in a double scaling limit,''
JHEP {\bf 9910}, 034 (1999).
[hep-th/9909110].
}

\lref\DiaconescuGU{
  D.~E.~Diaconescu and N.~Seiberg,
  ``The Coulomb branch of $(4,4)$ supersymmetric field theories in two-dimensions,''
JHEP {\bf 9707}, 001 (1997).
[hep-th/9707158].
}

\lref\AdamsSV{
  A.~Adams, J.~Polchinski and E.~Silverstein,
  ``Don't panic! Closed string tachyons in ALE space-times,''
JHEP {\bf 0110}, 029 (2001).
[hep-th/0108075].
}

\lref\VafaRA{
  C.~Vafa,
  ``Mirror symmetry and closed string tachyon condensation,''
in Shifman, M. (ed.) et al.: From fields to strings, vol. 3, 1828-1847.
[hep-th/0111051].
}

\lref\HarveyWM{
  J.~A.~Harvey, D.~Kutasov, E.~J.~Martinec and G.~W.~Moore,
  ``Localized tachyons and RG flows,''
[hep-th/0111154].
}
\lref\GutperleKI{
  M.~Gutperle, M.~Headrick, S.~Minwalla and V.~Schomerus,
  ``Space-time energy decreases under world sheet RG flow,''
JHEP {\bf 0301}, 073 (2003).
[hep-th/0211063].
}

\lref\MartinecWG{
  E.~J.~Martinec and G.~W.~Moore,
  ``On decay of K theory,''
[hep-th/0212059].
}

\lref\BakasEU{
  I.~Bakas,
  ``Renormalization group flows and continual Lie algebras,''
JHEP {\bf 0308}, 013 (2003).
[hep-th/0307154].
}
\lref\OliynykEY{
  T.~Oliynyk, V.~Suneeta and E.~Woolgar,
  ``Irreversibility of world-sheet renormalization group flow,''
Phys.\ Lett.\ B {\bf 610}, 115 (2005).
[hep-th/0410001].
}

\lref\BakasHN{
  I.~Bakas,
  ``Ricci flows and their integrability in two dimensions,''
Comptes Rendus Physique {\bf 6}, 175 (2005).
[hep-th/0410093].
}

\lref\future{
O.~Aharony, S.~Razamat, N.~Seiberg and B.~Willett, work in progress.}

\lref\afw{
O.~Aharony, A.~Feldman and B.~Willett, work in progress.}

\Title{
}
{\vbox{\centerline{The long flow to freedom}
\vskip0pt
}
}

\centerline{Ofer Aharony,$^a$ Shlomo S. Razamat,$^b$ Nathan Seiberg,$^c$ and Brian Willett$^{d}$}
\bigskip
\centerline{$^a$ {\it Department of Particle Physics and Astrophysics, Weizmann Institute of Science,}}
\centerline{{\it Rehovot 7610001, Israel}}
\centerline{$^b$ {\it Department of Physics, Technion, Haifa,  32000, Israel}}
\centerline{$^c$ {\it School of Natural Sciences, Institute for Advanced Study, Princeton, NJ 08540, USA}}
\centerline{$^d$ {\it Kavli Institute for Theoretical Physics,
University of California, Santa Barbara, CA 93106, USA}}

\vskip.2in \centerline{\bf Abstract}

\noindent
Two-dimensional field theories do not have a moduli space of vacua.  Instead, it is common that their low-energy behavior is a sigma model with a target space.  When this target space is compact its renormalization group flow is standard.  When it is non-compact the continuous spectrum of operators can change the qualitative behavior.  Here we discuss two-dimensional gauge theories with $\cN=(2,2)$ supersymmetry. We focus on two specific theories, for which we argue that they flow to free chiral multiplets at low energies: the $U(1)$ gauge theory with one flavor (two chiral superfields with charges plus and minus one) and a non-zero Fayet-Iliopoulos term, and pure $SU(N)$ gauge theories. We argue that the renormalization group flow of these theories has an interesting order of limits issue.  Holding the position on the target space fixed, the space flattens out under the renormalization group. On the other hand, if we first go to infinity on the target space and then perform the renormalization group, we always have a non-trivial space, e.g.\ a cone with a deficit angle. We explain how to interpret low-energy dualities between theories with non-compact target spaces.
We expect a similar qualitative behavior also for other non-compact sigma models, even when they do not flow to free theories.


\vskip.2in
{\centerline
{\it Dedicated to John Schwarz on his 75th birthday}
}
\noindent


\Date{}


\newsec{Introduction and Summary}

One characteristic feature of supersymmetric field theories is the existence of flat directions in the classical potential, which we will denote by $\cM_0$.  It is often the case that these lead in the quantum theory to a moduli space $\cM$ of inequivalent vacua.  The study of this space is one of the main tools in the analysis of these theories. $\cM$ is typically non-compact, and because of asymptotic freedom its asymptotic behavior is similar to that of the classical space $\cM_0$.

This situation changes in two dimensions.  Here there is no moduli space of vacua\foot{In \future\ we will review and discuss situations in which the Hilbert space of the quantum two-dimensional theory splits into a direct sum of Hilbert spaces, each being a separate superselection sector.  But even in these cases we do not obtain a continuum of different low-energy theories.}.  Instead, the fluctuations of the quantum theory explore all of $\cM$ and it can be viewed as a target space of the low energy excitations. Moreover, the metric of $\cM$ contains operators that are classically marginal, and that are crucial for understanding the quantum behavior of the theory, unlike in higher dimensions where only the complex structure of $\cM$ is important.

This fact has many interesting consequences.  One of them is a puzzle associated with duality.  Consider first an IR duality between two different higher dimensional theories $A$ and $B$.  Classically, each of them has its own space of flat directions $\cM_{0A}$ and $\cM_{0B}$, which become in the quantum theory $\cM_{A}$ and $\cM_{B}$.  Since typically $\cM_{0A}$ and $\cM_{0B}$ have different asymptotic metrics, and these are the same as those of $\cM_{A}$ and $\cM_{B}$ respectively, the asymptotic behaviors of $\cM_A$ and $\cM_B$ are different.  This does not contradict the IR duality between them, because the statement about the duality is only about the IR behavior.  The metric at a generic point on these spaces flows to a trivial flat metric in the IR.  And the duality simply means that at least one region in $\cM_A$ is similar to one region in $\cM_B$.

This cannot be the case in two dimensions.  First, as we said above, the quantum theory explores the whole target space $\cM$ and we cannot simply focus on one region.  Second, unlike the situation in the higher dimensional theory, the deviation of the metric on $\cM$ from the flat metric can be relevant or irrelevant. Thus, the metric on $\cM(\tau)$ depends non-trivially on the renormalization group parameter $\tau$. However, operators that change the asymptotic metric are not normalizable.  Therefore, one might conclude that two different two-dimensional theories $A$ and $B$ with different asymptotic metrics in $\cM_{0A}$ and $\cM_{0B}$ cannot possibly be dual to each other.

One of the points of this paper is that this conclusion could be too fast.  The reasoning in the previous paragraph involves two different limits: the IR limit $\tau \to \infty$ and going to the asymptotic region of $\cM(\tau)$.  We will argue that often these two limits do not commute.  When we discuss IR dualities in two dimensions we must {\it take the IR limit $\tau \to \infty$ before we go to infinity in $\cM(\tau)$}.  Being careful about this order of limits will allow us to clarify the status of various dualities. For any finite renormalization group time, the metric at infinity can be different in IR-dual theories. However, the metric at fixed positions in the interior of the space, and correlation functions of operators that are localized there, converge to the same values in IR-dual theories as we go to low energies.

Starting with \WittenYC,  there is a huge body of work analyzing $\cN=(2,2)$ supersymmetric field theories and their dynamics.  In this paper we will discuss two simple examples of such theories, which are conjectured (based on the comparison of objects protected by supersymmetry) to flow to free theories at low energies.  More precisely, the gauge theory first flows to a non-linear sigma model with some metric $\cM_0$, and then flows to the sigma model on $\cM(\tau)$ as above.  Further evidence for these dualities comes from reducing three dimensional dualities on a circle, as will be discussed in \future. In section 2 we will discuss the Higgs branch of a $U(1)$ gauge theory with two chiral superfields $Q$ and $\tilde Q$ with charges $\pm 1$. In section 3 we will discuss the pure gauge $SU(N)$ theories.  We expect the same general description of the renormalization group flows of non-compact sigma models, and of IR dualities between theories with non-compact moduli spaces, to arise also in interacting examples, such as the gauged linear sigma model for the non-compact conifold. Closely related behavior occurs also in the gauged linear sigma model for the two dimensional black hole \HoriAX. Some additional examples will be analyzed in \afw, including examples appearing in IR dualities discussed in \refs{\HoriDK\HoriPD\BeniniUI-\DoroudXW,\future}.

\ifig\uone{The Higgs branch of $\cN=(2,2)$ SUSY QED.  Figure $a$ depicts the classical Higgs branch $\cM_0(r=0)$ when the FI-parameter $r$ vanishes. Figure $b$ depicts the same space $\cM_0$ when the FI-term is non-zero.  It has the same asymptotic behavior as $\cM_0(r=0)$, but it is smooth. Figure $c$ shows the effect of the renormalization group flow on $\cM(\tau)$.  The asymptotic behavior of the space is not modified, but the metric is flattened out.  Hence the IR-dual theory is free.}
{\epsfxsize=5in
\epsfbox{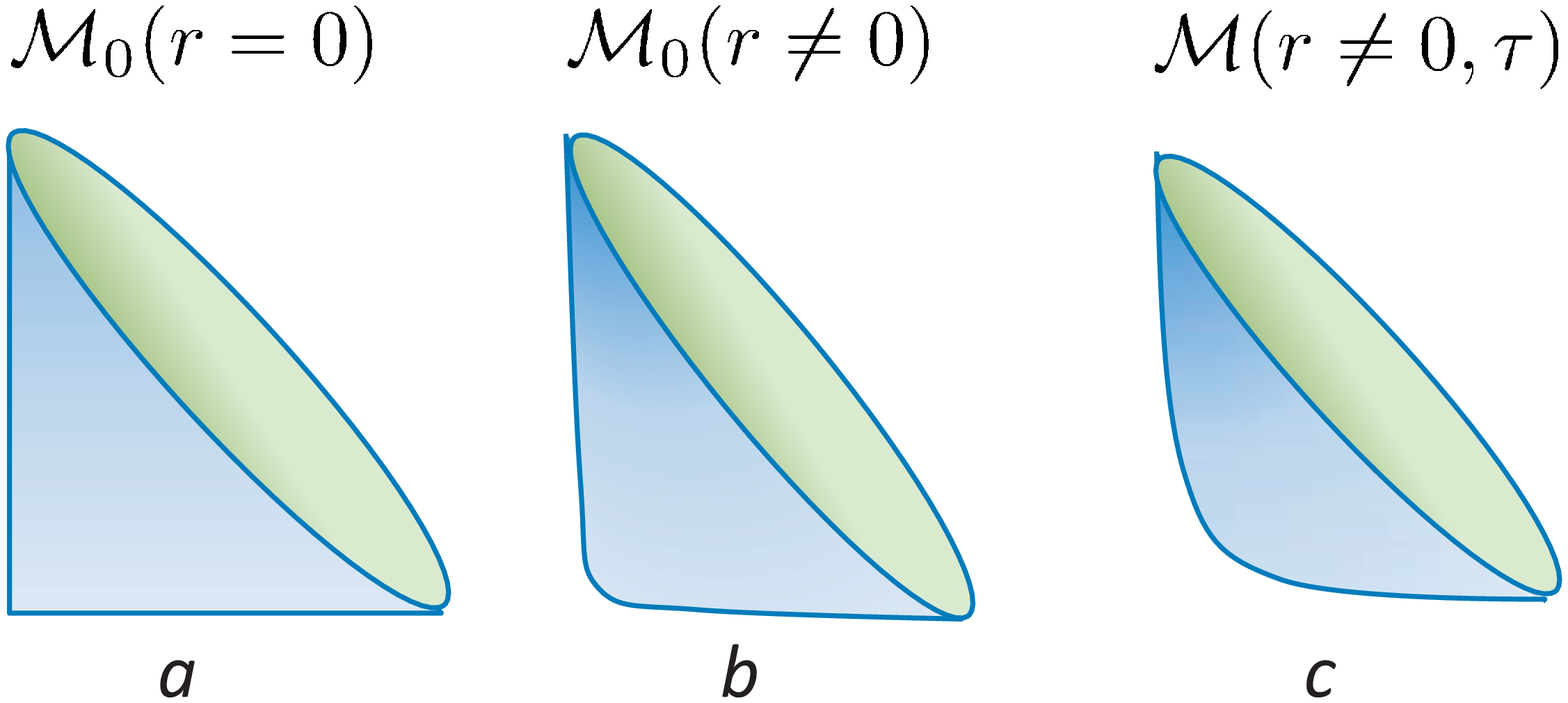}
}
\ifig\sutwo{The moduli space of $\cN=(2,2)$ SUSY Yang-Mills theory with gauge group $SU(2)$.  Figure $a$ depicts the classical Coulomb branch $\cM_0$.  We conjecture that in the quantum theory it is smoothed out by the renormalization group flow, and $\cM(\tau)$ is depicted in Figure $b$.  The IR dual theory is free.}
{\epsfxsize=5in
\epsfbox{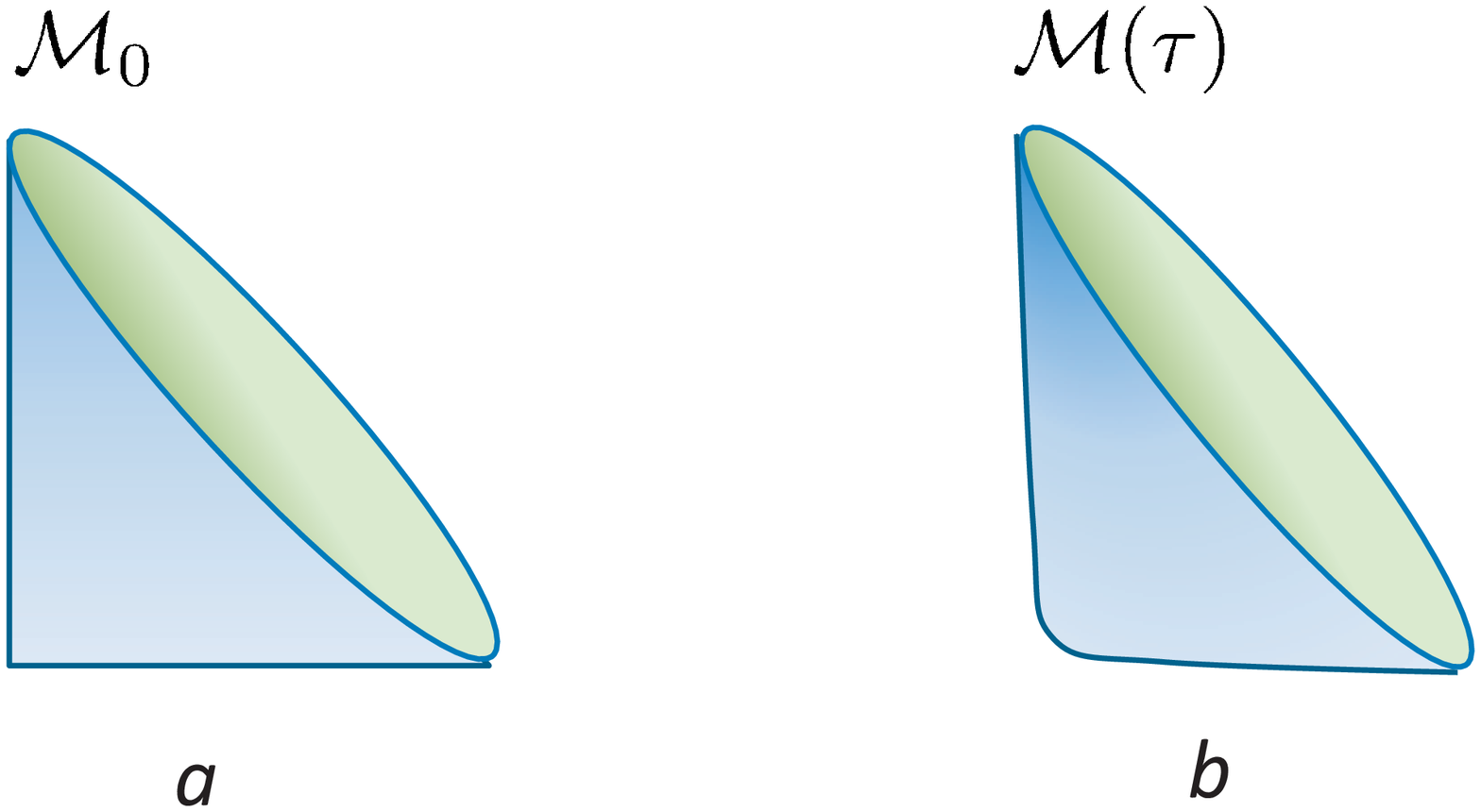}
}
The Abelian theory we discuss in section 2 can have a Fayet-Iliopoulos (FI) parameter $r$, and we will mostly be interested in non-zero $r$.  In \uone(a) we see the classical $\cM_0$ with $r=0$, which is the cone $\C/\Z_2$.  (We suppress the Coulomb branch, which emanates from the singular point.)  In \uone(b) we see $\cM_0$ for non-zero $r$, which is a capped cone.  We will argue that in the quantum theory $\cM(\tau)$ with non-zero $r$ is flattened out as in \uone(c).  The curvature is pushed to infinity and the space becomes flat, without changing the asymptotic behavior.  This way for any finite renormalization group (RG) time $\tau$ $\cM(\tau)$ has the same asymptotic behavior as $\cM_0$, but as $\tau \to \infty$ $\cM(\tau)$ becomes flat.  This makes it possible for this theory to be dual to a free field theory, whose asymptotic target space is $\C$ rather than $\C/\Z_2$. This is a special case of general IR dualities for $U(N)$ gauge groups, that were tested by the comparison of supersymmetric partition functions in \refs{\BeniniMIA,\BeniniXPA}.

In section 3 we discuss the pure $SU(N)$ gauge theory.  In the simplest case of $SU(2)$ its classical Coulomb branch is again a cone $\cM_0=\C/\Z_2$ (see \sutwo(a)).  We argue that its singularity is smoothed out by the renormalization group flow to \sutwo(b), such that it is dual (in the sense discussed above) to a free field theory, whose target space is $\C$. Similarly we argue that for all $N$ the theory is dual to $(N-1)$ free chiral multiplets.  A similar claim was made in \HoriDK\ for $SU(N)$ gauge theories with $N+1$ chiral multiplets in the fundamental representation.

Similar renormalization group flows in non-linear sigma models were analyzed from a different angle in \AdamsSV\ (some interesting followup papers are \refs{\HarveyWM\GutperleKI\MartinecWG-\VafaRA}). 
They were mostly interested in the use of these theories as worldsheet actions in string theory and in their space-time implications. One-loop renormalization group flows in non-compact non-linear sigma models were also analyzed in \refs{\BakasEU\OliynykEY-\BakasHN}. In this paper we are only interested in the two dimensional field theory and in the interpretation of low-energy dualities between gauge theories, and we focus on the change in the metric after long renormalization group time.

\newsec{$U(1)$ with one flavor}

Our main example will be a $U(1)$ gauge theory with one flavor, \ie, a pair of chiral multiplets with charges $\pm 1$.  The action of this theory can be written in superspace as:
\eqn\uoneaction{ S = \int d^2 x \int d^4 \theta  \left(\frac{1}{e^2} |\Sigma|^2  + Q^\dagger e^V Q+ \tilde{Q}^\dagger e^{-V} \tilde{Q} - t V\right). }
Here $e$ is the gauge coupling, with dimension one, and $t =r+i \theta$ is the complexified Fayet-Iliopoulos (FI) parameter, containing the ordinary FI parameter, $r$, and the theta angle, $\theta$. In the gauge theory $t$, which is classically marginal, does not flow under the RG. However, the low-energy theory has many other classically marginal operators, and we will see that they do flow non-trivially under the RG.

In addition to R-symmetries, this theory has a single $U(1)_a$ global symmetry, under which both chiral multiplets transform with charge $1$.  One can weakly gauge this symmetry and turn on a background value for the complex scalar in the gauge multiplet; such a ``twisted mass'' deforms the theory to a massive theory with a single supersymmetric vacuum for any value of the FI parameter.  However, for most of what follows, we will work at vanishing twisted mass, and then we expect the theory defined by the action \uoneaction\ to flow to a superconformal field theory at low energies.

At energies below the photon mass, proportional to the gauge coupling $e$, the photon can be integrated out and the theory can be described in terms of a non-linear sigma model, whose target space $\cM_0$ can be obtained by solving the $D$-term equations modulo gauge symmetry:
\eqn\modulispace{ \cM_0 = \{ |Q|^2 - |\tilde{Q}|^2 -r = 0 \} /U(1)_{gauge}.
}
This space is topologically a complex plane, with a natural gauge-invariant chiral coordinate $M \equiv Q \tilde{Q}$. The metric on this space does not receive corrections depending on $e$, but as a non-linear sigma model $\cM(\tau)$ flows non-trivially.

We will describe the geometry of this space in more detail below.  For $r$ non-zero, it is smooth, while for $r=0$ it has a conical singularity at $M=0$, where the photon is classically massless.  If the theta angle is also tuned appropriately,\foot{Specifically, this singularity occurs at $\theta=\pi$, rather than $\theta=0$ as one might naively expect, because the effective theta angle at low energies gets a contribution of $\pi$ relative to the bare theta angle after integrating out the chiral multiplets.} we expect a singularity in the CFT, associated to the presence of a Coulomb branch emanating from the point $M=0$, on which $\Sigma$ takes non-zero values. This theory is expected to have a continuous spectrum related to a ``throat'' that develops near the origin and connects (for finite RG time) the Higgs branch to the Coulomb branch, and we will not discuss it further here.

In the remainder of this section we will provide evidence that this theory, for non-zero $r$, is IR-dual to a free chiral multiplet, which can be identified with $M=Q \tilde{Q}$.  This duality is a special case of a general $U(N)$ duality proposed in \BeniniMIA. It is not expected to hold for the special value $t=i\pi$, for which we do not expect the RG flow to lift the continuum of states living near the origin. But we conjecture that it holds whenever $r \neq 0$.

The supersymmetric partition functions of the $U(1)$ gauge theory and the free theory, on both $\S^2$ and $T^2$, have been shown to agree  in \refs{\BeniniMIA\BeniniXPA-\BeniniNDA}.
The precise relations one obtains from equation (2.45) in \BeniniMIA, and from section 4.6.1 of \BeniniXPA, are :
\eqn\pfduality{
\cZ^{\S^2}_{U(1) \; N_f=1  }(\mu,t,\bar{\mu},\bar{t}) = e^{ {\tilde{W}}_{BG}(\mu,t) -c.c.} \cZ_{chiral}^{\S^2}(2 \mu,2 \bar{\mu}), \;\;\;\;\;  \cZ^{T^2}_{U(1) \; N_f=1} (q,y,\nu) = \cZ_{chiral}^{T^2}(q,y,\nu^2), }
for appropriately defined partition functions that can be computed by localization \refs{\BeniniUI,\DoroudXW,\BeniniNDA}.  In the $T^2$ partition function, $q=e^{2 \pi i\tau}$ is the modular parameter, and $y$ and $\nu$ are fugacities for the $U(1)_R$ and $U(1)_a$ symmetries, respectively.
The $\S^2$ partition function agrees up to a certain contact term, namely, a twisted superpotential depending on the background fields:
\eqn\contacta{
{\tilde{W}}_{BG}(\mu,t) = -2 \mu \log \bigg(2 \cosh \left(\frac{t}{2}\right) \bigg),
}
where $\mu$ is the scalar in the background gauge multiplet coupled to the $U(1)_a$ flavor symmetry.  Note that this term diverges at $t= i \pi$, related to the extra light degrees of freedom there, and indicating that the duality fails at this point.

\subsec{ The K\"ahler potential and its one-loop renormalization group flow}

The partition function checks of this duality are quite non-trivial, but they suffer from the basic shortcoming that they are only sensitive to protected information in the theory.  For the theories to be precisely dual, they must agree not only in their protected data, but also in unprotected data that affects the low-energy effective action, such as the K\"ahler potential for $M$.
Precisely because such data is not protected, it is difficult to check that it matches across a putative duality.  Moreover, as we mentioned in the introduction and will explain in more detail below, the fact that the asymptotic behavior of the target space seems different poses a challenge to the duality.
However, at least in some limit of parameters, we will find evidence that the K\"ahler potential does match across the duality.

Recall that at energies far below the gauge coupling $e$, the gauge field (which is massive on the Higgs branch for generic values of $r$) can be integrated out to find an approximate description as a non-linear sigma model into the space \modulispace.  Let us start by computing the tree-level K\"ahler potential for the gauge theory.
In the limit $e^2 \rightarrow \infty$, the gauge kinetic term can be ignored, and the K\"ahler potential is
given by:
\eqn\uonekahler{
 K = Q^\dagger e^V Q+ \tilde{Q}^\dagger e^{-V} \tilde{Q} - r V.
}
The $D$-term constraint on the chiral multiplets is:
\eqn\uonedterm{
Q^\dagger e^V Q - \tilde{Q}^\dagger e^{-V} \tilde{Q} - r = 0.
}
To solve this, let us write $X=Q^\dagger e^V Q$ and $\tilde{X} = \tilde{Q}^\dagger e^{-V} \tilde{Q}$, and note:
\eqn\uonesol{
X \tilde{X} = |M|^2, \;\;\; X - \tilde{X} - r= 0. }
These can be solved for $X$ and $\tilde{X}$ in terms of $|M|^2$ and $r$, and we find:
\eqn\uonekahlersol{
K= \sqrt{r^2+4 |M|^2 }- |r| \log (|r| + \sqrt{r^2+ 4 |M|^2}).
}
This leads to a tree-level metric:
\eqn\uonetreelevelmetric{ ds^2 = \frac{1}{\sqrt{r^2 + 4 |M|^2}} dM d\bar{M}. }

Let us first note a few properties of this metric on $\cM_0$, depicted in figures \uone(a) and \uone(b).  For $|M|\gg |r|$, the metric behaves as:
\eqn\uonetreelevelmetriclargeM{ ds^2 \approx \frac{1}{2 |M|} dM d\bar{M}, }
which is flat, but with a deficit angle of $\pi$ at infinity.  As $r \rightarrow 0$, the metric approaches that of a cone, $\C/\Z_2$, which is flat everywhere except for a conical singularity at $M=0$.  For finite $r$ the tip of the cone is smoothed, with curvature of order $1/|r|$ (going to zero as $M\to \infty$).  As $|r| \rightarrow \infty$, the space becomes locally flat.

Note from \uonetreelevelmetriclargeM\ that the asymptotic behavior of the metric is very different from that of the putative dual, which is given by the flat metric, $ds^2=dM d\bar{M}$.  On the other hand, \uonetreelevelmetriclargeM\ is asymptotically flat, so one does not expect it to undergo renormalization group flow.  This leads to an apparent contradiction with the proposed duality.

To resolve it, we study the RG flow of the metric.  This is governed by the equation:
\eqn\rgflow{
\partial_\tau g_{i \bar{j}} = -\frac{1}{2 \pi} R_{i \bar{j}} + O(R^2), }
where $R_{i \bar{j}}$ is the Ricci tensor, and $\tau$ is the RG time.  We expect the first term of this equation to give a good approximation to the RG flow when the curvature is everywhere small.  This holds for \uonetreelevelmetric\ provided we take $|r|$ sufficiently large.  If we write $M=e^{u+i \theta}$, for $u \in \R$, then \uonetreelevelmetric\ takes the form:
\eqn\genmetric{ ds^2 = \Omega(u) (du^2 + d\theta^2),}
with:
\eqn\initialcondition{
\Omega(u) = \frac{1}{|r|} \frac{e^{2u}}{\sqrt{1 + \frac{4}{r^2} e^{2 u}}}.
}
Then the one-loop approximation to the RG equation \rgflow\ becomes:
\eqn\rgflowtwo{ \partial_\tau \Omega(u,\tau) = \frac{1}{4 \pi} \frac{\partial^2}{\partial u^2}\bigg( \log (\Omega(u,\tau)) \bigg),}
with initial condition given by \initialcondition.  Note that redefining $u \rightarrow u+\log (|r|/2)$ and $\tau \rightarrow |r| \tau$ eliminates $r$ from this equation.  Thus the one-loop RG equation is essentially independent of $r$, and so we can set $r=2$ for simplicity. Corrections to \rgflowtwo\ are proportional to powers of $1/|r|$.

\ifig\fone{$\Omega/v^2$ as a function of $v$ for various RG times.  The bottom curve corresponds to the initial, tree-level metric, while the line at $\Omega/v^2=1$ corresponds to the flat metric.}
{\epsfxsize=4in
\epsfbox{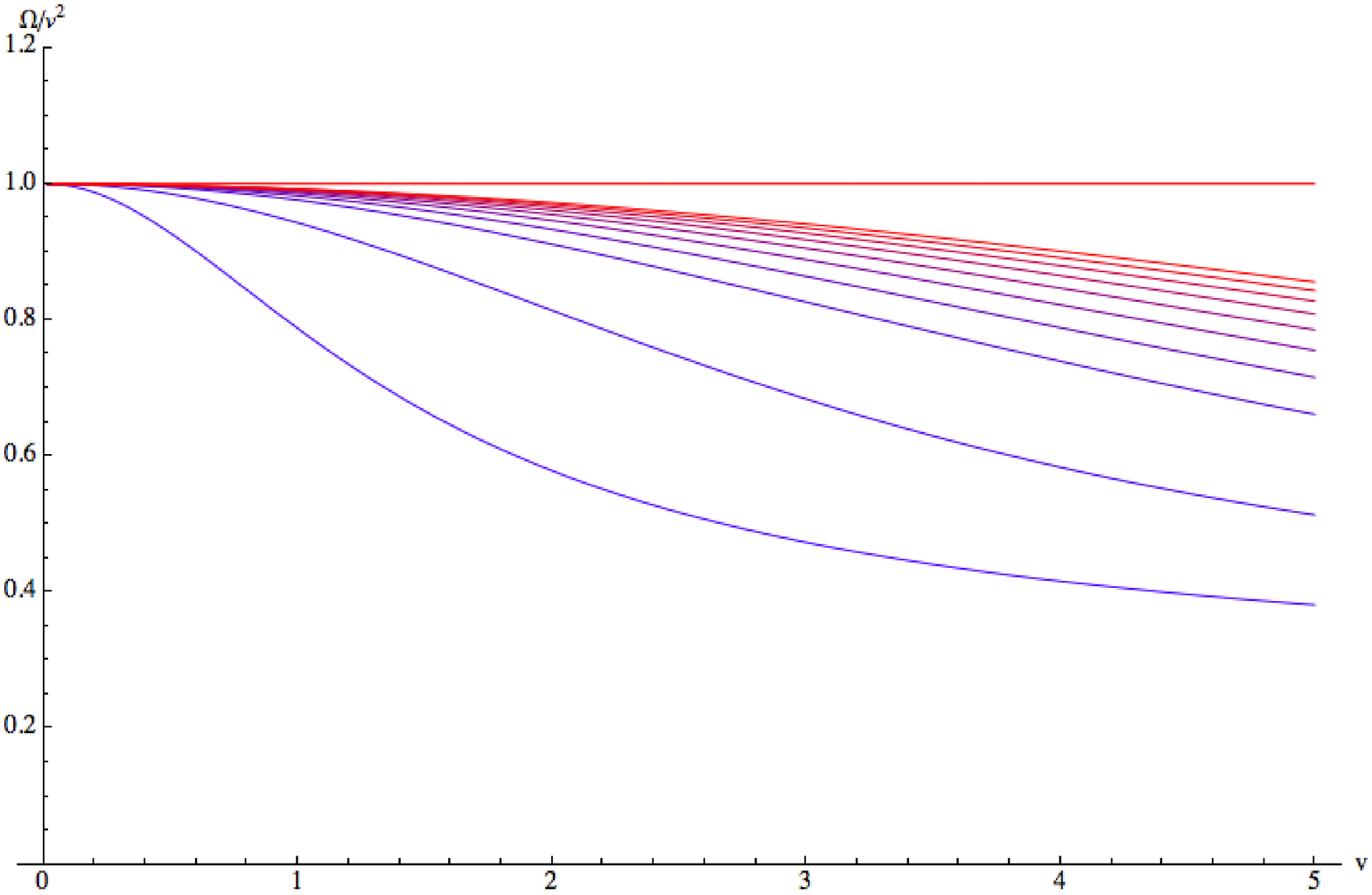}
}

\ifig\ftwo{$\sqrt{g} R$ as a function of $v$ for various RG times.  The tallest peak corresponds to the initial metric, whose curvature is concentrated near the tip of the cone at $v=0$.  As RG time increases, this moves to positive $v$ and spreads outward.}
{\epsfxsize=4in
\epsfbox{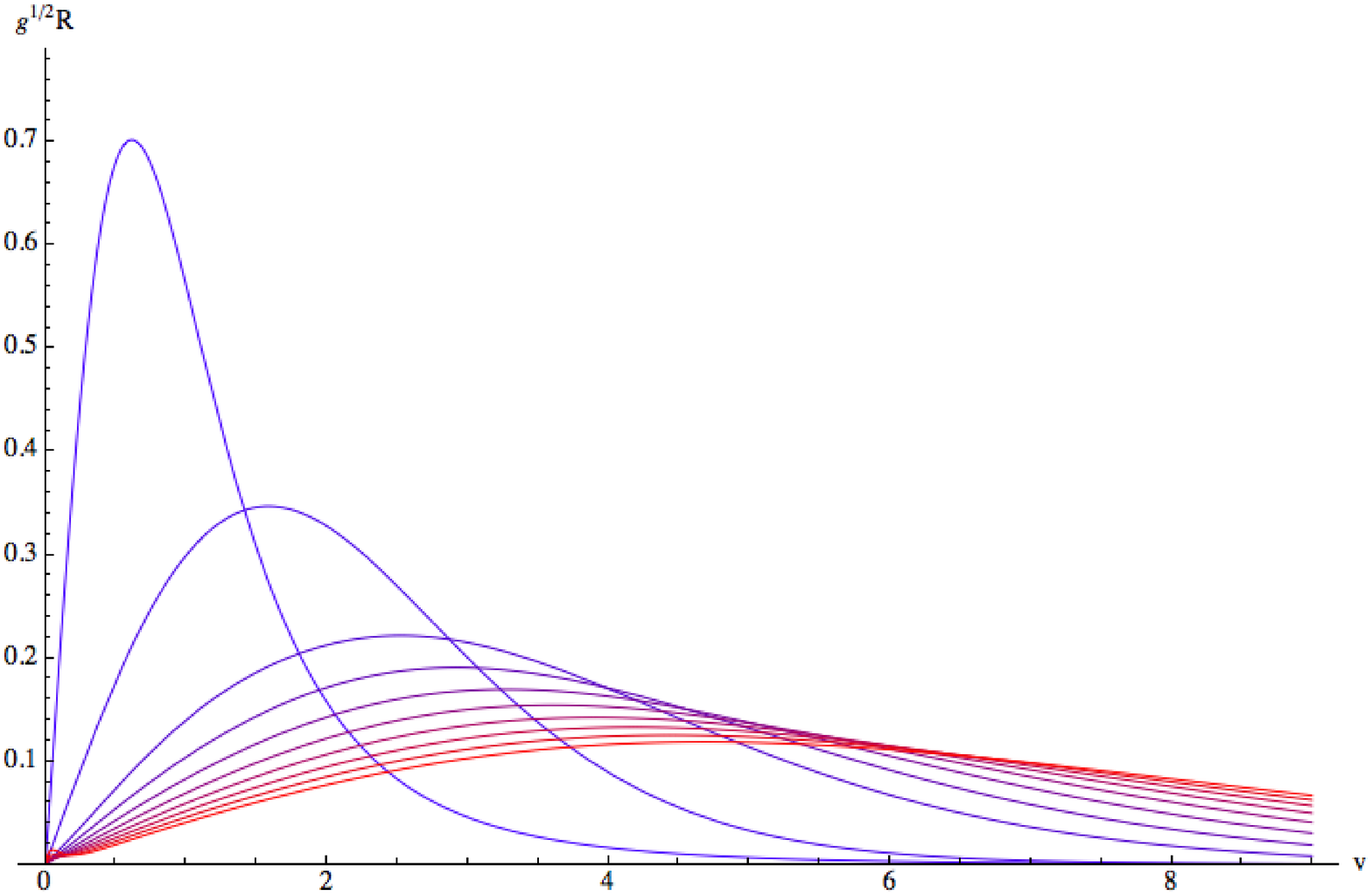}
}

We solve \rgflowtwo\ numerically using Mathematica  The results are shown in \fone\ and \ftwo, for RG times up to $\tau=300$.  Here we have used an RG-time-dependent coordinate transformation to place the metric in the form:
\eqn\niceform{
 ds^2 = dv^2 + \Omega(v) d\theta^2.}
In \fone\ we plot the ratio $\Omega/v^2$.  The asymptotic behavior of the initial metric is $\Omega/v^2 \approx \frac{1}{4}$, and as RG time increases we see that the behavior at any fixed value of $v$ approaches that of the flat metric, with $\Omega/v^2 =1$.  In \ftwo\ we plot $\sqrt{g} R$, where $R$ is the Ricci scalar.  Here we see that the non-zero curvature is moving outwards from the origin, and spreading out\foot{The standard intuition that positive curvature flows to larger values and negative curvature flows to smaller values applies in symmetric spaces.  This is not the case here and indeed our positive curvature near the origin flows to smaller values.}.  Note that the integral of this quantity over $v$ is a topological quantity, and thus it is conserved by the RG flow.

These results indicate that the space near the tip of the cone smooths out with RG time.  This flattening modifies the region near the tip to look like flat space.  Although the asymptotic behavior is unchanged by the RG flow, the flat region will continue to spread outward, and for any fixed position, the metric near this point will eventually look like the flat metric, for sufficiently long RG times.  A similar behavior was observed in \AdamsSV\ in the context of theories with $\C/\Z_n$ target space.  As argued there, the smoothing of the tip of the cone is consistent with the fact that the RG flow tends to reduce the volume of the target space in the region interior to a given radius.

In this sense, the theory flows in the infrared to the theory of a free chiral multiplet, with target space $\cM(\tau=\infty)$ which is the flat complex plane. The IR duality is the claim that if we first flow to the IR, and then compute correlation functions of operators that are {\it localized} in the target space, then they will be the same for the gauge theory and for the free theory.

The numerical results above give a good description of the RG flow of the metric only when $|r| \gg 1$, so that the one-loop approximation can be trusted.  Nevertheless, we conjecture that this qualitative behavior persists for any non-zero $r$, and all such theories flow to the free theory.

\newsec{Pure $SU(N)$ gauge theories}

Pure $SU(N)$ $\cN=(2,2)$ gauge theories have a complex scalar $\Sigma$ in the adjoint representation,
which can be viewed as the bottom component of a twisted chiral multiplet. They have a classical Coulomb branch of vacua of
dimension $N-1$, where the gauge group is broken to $U(1)^{N-1}$. It may be parameterized by the eigenvalues
$\Sigma_i$ of $\Sigma$, satisfying $\sum_{i=1}^N \Sigma_i = 0$, which are defined up to permutations (the Weyl group of $SU(N)$).
The effective twisted superpotential on this
Coulomb branch vanishes, so one expects it to remain also in the
quantum theory.

One can obtain an effective theory on the Coulomb branch by integrating
out the W-bosons. It is natural to have a kinetic term for $\Sigma$ proportional to $1/g_{YM}^2$, and then the classical W-boson masses are given by $|\Sigma_i-\Sigma_j|$. Classically the metric on the Coulomb branch is
conical, with singularities where W-bosons become massless; for instance for $N=2$ it is
$\C/\Z_2$.
Because of the Weyl group identifications on the $\Sigma_i$ coordinates, it is convenient to use gauge-invariant coordinates
\eqn\ydef{Y_n \equiv {\rm tr}(\Sigma^n) = \sum_{i=1}^N
\Sigma_i^n}
for $n=2,3,\cdots,N$. The classical
metric is flat and canonical in the $\Sigma_i$ variables
\eqn\classmet{{1\over g_{YM}^2} \sum_{i=1}^N |d\Sigma_i|^2 ~.}
This can be translated into a metric for the $Y_n$ variables using \ydef.

Quantum corrections modify \classmet\ by a power series in $g_{YM}^2 / \Sigma^2$ (where $\Sigma$ denotes masses of W-bosons); in canonically normalized dimensionless variables ${\tilde \Sigma}_i = \Sigma_i / g_{YM}$, they give a power series in $1/{\tilde \Sigma}^2$.
The low-energy metric obtained by integrating out the W-bosons is canonical (flat) in the $\Sigma_i$ coordinates far from the singularities where W-bosons are massless, but receives large perturbative and non-perturbative corrections there. So we have at energies below $g_{YM}$ a sigma model with a metric which is a function of ${\tilde \Sigma}_i$, or equivalently of the single-valued $Y_n$, which we do not know how to compute exactly, though we know its form when all $|\Sigma_i-\Sigma_j|$ are large.

We conjecture that :
\item {\bf A.}
The quantum corrections smooth out the classical singularities, such that the metric  becomes smooth everywhere.
\item {\bf B.}
The resulting sigma model flows under the renormalization group to a flat metric on $\C^{N-1}$ with a canonical form in terms of the gauge-invariant variables $Y_n$. Namely, we conjecture that the pure $SU(N)$ gauge theory is IR-dual to free fields $Y_n$, in the sense discussed above. Note that in terms of these variables the original metric is  very different from the original canonical metric in terms of $\Sigma_i$ even far from the singularities, so this flow requires a change in the asymptotic form of the metric, as in the previous section. For the pure $SU(2)$ theory, the conjectured change in the asymptotic form of the metric happens to be the same as the one discussed in the previous section.

\bigskip

Since we do not know the original metric of the gauge theory with the quantum corrections, our tests of this conjecture are weaker than in the previous section, and we cannot directly study the RG flow.

One piece of evidence for this conjecture is that the
elliptic genus of the pure $SU(N)$ gauge theory is the same
as that of the theory of $(N-1)$ free twisted chiral superfields
$Y_n$.  From \BeniniXPA, the elliptic genus of the pure $SU(N)$ theory is given by:
\eqn\sunellgen{
\cZ^{T^2}_{SU(N)}(q,y) = \frac{\theta(y^{-1};q)}{\theta(y^{-N};q)}. }
To compare this to the elliptic genus of the $N-1$ free twisted chirals $Y_n$, $n=2,...,N$, note that a free twisted chiral of R-charge $r$ contributes:
\eqn\ttwotwistedchiral{ \cZ^{T^2}_{twisted \; chiral}(q,y) = \frac{\theta(y^{1-r/2};q)}{\theta(y^{-r/2};q) }.}
Since ${\rm tr}(\Sigma^n)$ has R-charge $2n$, the collection of these free fields has elliptic genus:
\eqn\sundual{
\cZ^{T^2}_{\otimes_n tr(\Sigma^n)}(q,y) = \prod_{n=2}^N \frac{\theta(y^{1-n};q)}{\theta(y^{-n};q)}  = \frac{\theta(y^{-1};q)}{\theta(y^{-N};q)} ,
}
agreeing with \sunellgen.  Note that the IR theory has an enhanced $U(N-1)$ flavor symmetry, but since this is not visible in the UV we cannot turn on the corresponding fugacities in the elliptic genus.

For this theory the $\S^2$ partition function diverges (there are no twisted chiral parameters that can be used to regularize it), so we cannot use it to test the duality.

Another important prediction of this scenario, which passes many
consistency checks, is the following. Non-Abelian gauge theories
with a generic spectrum of massive chiral multiplets have an
effective twisted superpotential ${\tilde W}$ on their Coulomb branch, which can
be computed exactly at one-loop. The extrema
of this superpotential are expected to correspond to supersymmetric
vacua. However, sometimes extrema appear at points where some of the
$\Sigma_i$ are equal, so that some non-Abelian gauge group is classically unbroken
(with no massless matter fields); this gauge
theory is strongly coupled so the naive analysis based on the effective superpotential cannot
be trusted.  However, there is a well-defined question of whether there
is a supersymmetric vacuum at these points or not. Evidence from the
 index and other considerations suggests (see, for instance, \HoriDK) that there is actually
no supersymmetric vacuum there.

In the scenario above, the effective twisted superpotential near the points
of enhanced $SU({\tilde N})$ is naturally written in the ${\tilde Y}_n$ variables of this effective pure $SU({\tilde N})$ theory, which has a smooth metric there. Generically, when expanded near these points,
the effective superpotential $\tilde W$ will contain terms linear in the ${\tilde Y}_n$
(plus higher order terms). This implies that there should not
be any supersymmetric vacuum at the point ${\tilde Y}_2 = {\tilde Y}_3 = \cdots = 0$
of enhanced $SU({\tilde N})$, consistent with the known results.

\bigskip
\noindent{\bf Acknowledgments}

We would like to thank F.~Benini and K.~Hori for useful discussions.
The work of OA and SSR was supported in part  by the I-CORE program of the Planning and Budgeting Committee and the Israel Science Foundation (grant number 1937/12). The work of OA was supported in part by an Israel Science Foundation center for excellence grant, by the Minerva foundation with funding from the Federal German Ministry for Education and Research, by a Henri Gutwirth award from the Henri Gutwirth Fund for the Promotion of Research, and by the ISF within the ISF-UGC joint research program framework (grant no.\ 1200/14). OA is the Samuel Sebba Professorial Chair of Pure and Applied Physics.  SSR is  a Jacques Lewiner Career Advancement Chair fellow. The research of SSR was also supported by the Israel Science Foundation under grant no.\ 1696/15.  NS was supported in part by DOE grant DE-SC0009988. NS thanks the hospitality of the Weizmann Institute of Science during
the completion of this work.  BW was supported in part by the National Science Foundation under Grant No. NSF PHY11-25915.

\listrefs
\end